\documentclass[aps,showpacs,showkeys]{revtex4}

\usepackage{amssymb,amsmath}
\usepackage[pdftex]{graphicx}

\newcommand{\bd}[1]{\mbox{\boldmath$#1$}}

\newcommand{\pd}[1]{\ensuremath{[\![#1]\!]}}
\newcommand{\Pd}[1]{\ensuremath{\Bigl[\!\!\Bigl[#1\Bigr]\!\!\Bigr]}}
\newcommand{\PD}[1]{\ensuremath{\biggl[\!\!\biggl[#1\biggr]\!\!\biggr]}}

\begin{document}

\title{Phase description of oscillatory convection with a spatially translational mode}

\author{Yoji Kawamura}
\email[Corresponding author: ]{ykawamura@jamstec.go.jp}
\affiliation{Institute for Research on Earth Evolution,
Japan Agency for Marine-Earth Science and Technology, Yokohama 236-0001, Japan}
\affiliation{Department of Mathematical Science and Advanced Technology,
Japan Agency for Marine-Earth Science and Technology, Yokohama 236-0001, Japan}

\author{Hiroya Nakao}
\affiliation{Department of Mechanical and Environmental Informatics, Tokyo Institute of Technology, Tokyo 152-8552, Japan}

\date{April 22, 2014} 



\begin{abstract}
  We formulate a theory for the phase description of oscillatory convection
  in a cylindrical Hele-Shaw cell that is laterally periodic.
  This system possesses spatial translational symmetry in the lateral direction owing to the cylindrical shape
  as well as temporal translational symmetry.
  Oscillatory convection in this system is described by a limit-torus solution
  that possesses two phase modes; one is a spatial phase and the other is a temporal phase.
  The spatial and temporal phases indicate the ``position'' and ``oscillation'' of the convection, respectively.
  The theory developed in this paper can be considered as a phase reduction method
  for limit-torus solutions in infinite-dimensional dynamical systems,
  namely, limit-torus solutions to partial differential equations
  representing oscillatory convection with a spatially translational mode.
  We derive the phase sensitivity functions for spatial and temporal phases;
  these functions quantify the phase responses of the oscillatory convection
  to weak perturbations applied at each spatial point.
  Using the phase sensitivity functions,
  we characterize the spatiotemporal phase responses of oscillatory convection to weak spatial stimuli
  and analyze the spatiotemporal phase synchronization between weakly coupled systems of oscillatory convection.
\end{abstract}

\maketitle

\begin{quotation}
{ \noindent
  {\bf Highlights:} \\
  We develop a phase reduction theory for oscillatory convection with a spatial mode.   \\
  The theory can be considered as a phase description method for limit-torus solutions. \\
  We derive phase sensitivity functions for spatial and temporal phases of convection.  \\
  We can quantify spatiotemporal phase responses of convection to weak perturbations.   \\
  We can analyze spatiotemporal phase synchronization between weakly coupled systems.   \\
}
\end{quotation}


\section{Introduction} \label{sec:1}


Nature provides abundant examples of rhythmic systems and synchronization phenomena
\cite{ref:winfree80,ref:kuramoto84,ref:pikovsky01,ref:strogatz03,ref:manrubia04}.
Each rhythmic system is typically described by an ordinary differential equation that possesses a limit-cycle solution.
The phase reduction method for ordinary limit-cycle oscillators
has been well established and successfully applied to analyze the synchronization properties of oscillators
\cite{ref:winfree80,ref:kuramoto84,ref:pikovsky01,ref:hoppensteadt97,ref:izhikevich07,ref:ermentrout10,ref:ermentrout96,ref:brown04}.
There also exist rhythmic spatiotemporal patterns described by limit-cycle solutions to partial differential equations
\cite{ref:manneville90,ref:mori97,ref:cross93,ref:cross09,ref:mikhailov06,ref:mikhailov13,ref:lappa10,ref:lappa12}.

We recently developed a phase description method for limit-cycle solutions to the following partial differential equations:
the nonlinear Fokker-Planck equations
that represent the collective dynamics of globally coupled noisy dynamical elements
\cite{ref:kawamura11},
the fluid equations
that represent the dynamics of the temperature field in ordinary Hele-Shaw cells
\cite{ref:kawamura13},
and the reaction-diffusion equations
that represent rhythmic spatiotemporal patterns in chemical and biological systems
\cite{ref:nakao12}.
However, there are also examples of spatiotemporal rhythms
in systems that further possess spatial translational symmetry;
these spatiotemporal rhythms cannot be described by limit-cycle solutions.

For example, rotating annuli and spheres possess continuously rotational symmetry,
i.e., continuously translational symmetry in the rotating direction
\cite{ref:ghil87,ref:boubnov95,ref:lappa10,ref:lappa12}.
Consequently, the emergence of spatiotemporal rhythms in such systems
brings up two phase modes, i.e., a spatial phase and a temporal phase.
Such spatiotemporal rhythms are described by limit-torus solutions.
Synchronization of spatiotemporal rhythms with two phase modes has been experimentally investigated
using systems of rotating fluid annuli that exhibit traveling and oscillating convection,
which is analogous to atmospheric circulation~\cite{ref:read09,ref:read10}.
Therefore, a phase description method for limit-torus solutions to partial differential equations is desirable.

In this paper, as the first step,
we consider oscillatory convection in a cylindrical Hele-Shaw cell that is laterally periodic.
An ordinary Hele-Shaw cell is a rectangular cavity where the gap between two vertical walls
is much smaller than the extent of the other two spatial dimensions,
and the fluid in the cavity exhibits oscillatory convection under the appropriate parameter conditions
(see Refs.~\cite{ref:bernardini04,ref:nield06} and references therein).
The cylindrical Hele-Shaw cell is a cylindrical version of the ordinary Hele-Shaw cell
that possesses spatial translational symmetry in the lateral direction owing to the cylindrical shape.
Oscillatory convection in the cylindrical Hele-Shaw cell
is therefore described by a limit-torus solution that possesses both spatial and temporal phases.

Here, we formulate a theory for the phase description of oscillatory convection in the cylindrical Hele-Shaw cell.
The theory can be considered as a phase reduction method for limit-torus solutions to partial differential equations.
The theory can also be considered as a generalization of our phase description method
for limit-cycle solutions to partial differential equations
such as the nonlinear Fokker-Planck equations~\cite{ref:kawamura11},
fluid equations~\cite{ref:kawamura13},
and reaction-diffusion equations~\cite{ref:nakao12}.
The phase reduction method for limit-torus solutions
enables us to describe the dynamics of the oscillatory convection by two phases (i.e., spatial and temporal phases),
and facilitates theoretical analysis of the spatiotemporal phase synchronization properties of the oscillatory convection.
On the basis of phase reduction,
we characterize the spatiotemporal phase responses of oscillatory convection to weak impulses
and analyze the spatiotemporal phase synchronization between weakly coupled systems exhibiting oscillatory convection.

This paper is organized as follows.
In Sec.~\ref{sec:2},
we formulate a theory for the phase description
of oscillatory convection with a spatially translational mode;
supplemental information of the theory is given in App.~\ref{sec:A} and App.~\ref{sec:B}.
In Sec.~\ref{sec:3},
we illustrate the theory using a numerical analysis of the oscillatory convection.
In Sec.~\ref{sec:4},
we make a comparison between the theory and direct numerical simulations.
Concluding remarks are given in Sec.~\ref{sec:5}.

\section{Phase description of oscillatory convection} \label{sec:2}

In this section,
we formulate a theory for the phase description of oscillatory convection
in a cylindrical Hele-Shaw cell that is laterally periodic.
The theory can be considered as an extension of our phase description method
for oscillatory convection in the ordinary Hele-Shaw cell~\cite{ref:kawamura13}
to that in the cylindrical Hele-Shaw cell.

\subsection{Dimensionless form of the governing equations}

\begin{figure*}
  \begin{center}
    \includegraphics[width=0.5\hsize,clip]{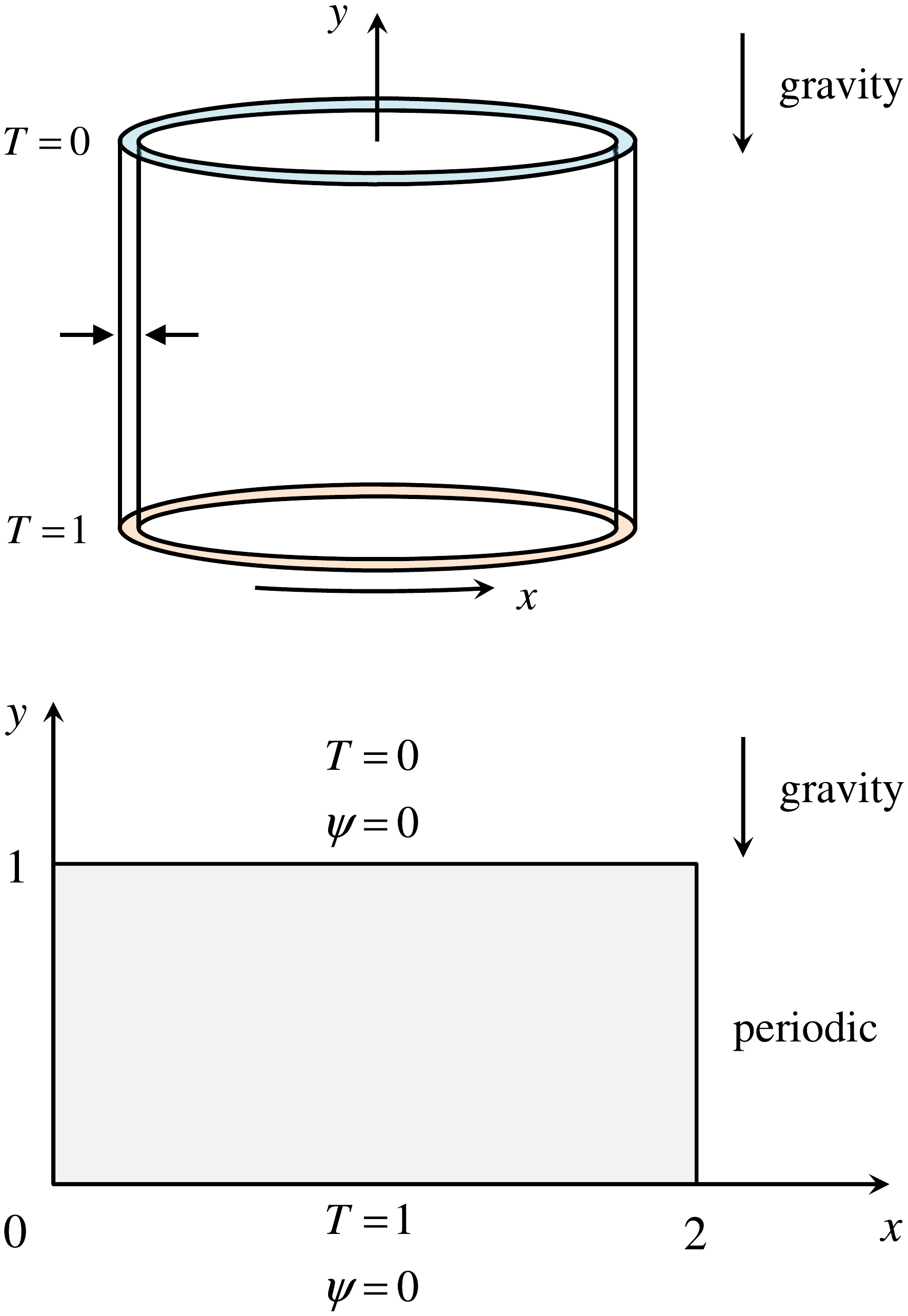}
  \end{center}
  \caption{(Color online)
    Schematic diagram of the cylindrical Hele-Shaw cell that is laterally periodic.
    Curvature effects due to the cylindrical shape are assumed to be negligible.
    The temperature at the bottom ($y = 0$) is higher than at the top ($y = 1$).
  }
  \label{fig:1}
\end{figure*}

The dynamics of the temperature field $T(x, y, t)$ in the cylindrical Hele-Shaw cell
is described by the following dimensionless form
(see Ref.~\cite{ref:bernardini04} and references therein):
\begin{align}
  \frac{\partial}{\partial t} T(x, y, t)
  = \nabla^2 T + J(\psi, T).
  \label{eq:T}
\end{align}
The Laplacian and Jacobian are respectively given by
\begin{align}
  \nabla^2 T
  &= \left( \frac{\partial^2}{\partial x^2}
  + \frac{\partial^2}{\partial y^2} \right) T,
  \\
  J(\psi, T)
  &= \frac{\partial \psi}{\partial x} \frac{\partial T}{\partial y}
  - \frac{\partial \psi}{\partial y} \frac{\partial T}{\partial x},
\end{align}
where we assumed that the curvature effects due to the cylindrical shape are negligible
(see Refs.~\cite{ref:zhao90,ref:miranda02} for curvature effects,
although the subject of these references is not thermal convection but viscous fingering).
The first and second terms on the right-hand side of Eq.~(\ref{eq:T})
represent diffusion and advection, respectively.
The stream function $\psi(x, y, t)$ is determined from the temperature field $T(x, y, t)$ as follows:
\begin{align}
  \nabla^2 \psi(x, y, t) = -{\rm Ra} \frac{\partial T}{\partial x},
  \label{eq:P_T}
\end{align}
where the Rayleigh number is denoted by ${\rm Ra}$.
The stream function $\psi(x, y, t)$ also gives the fluid velocity field $\bd{v}(x, y, t)$, i.e.,
\begin{align}
  \bd{v}(x, y, t)
  = \left( \frac{\partial \psi}{\partial y}, \, -\frac{\partial \psi}{\partial x} \right).
\end{align}
Figure~\ref{fig:1} shows a schematic diagram of the cylindrical Hele-Shaw cell.
The system is defined in the following rectangular region: $x \in [0, 2]$ and $y \in [0, 1]$.
Because this Hele-Shaw cell has a cylindrical shape,
the system possesses a $2$-periodicity with respect to $x$.
The boundary conditions for the temperature field $T(x, y, t)$ are given by
\begin{align}
  & T(x + 2, y, t) = T(x, y, t),
  \label{eq:bcTx} \\
  & \Bigl. T(x, y, t) \Bigr|_{y = 0} = 1, \qquad \Bigl. T(x, y, t) \Bigr|_{y = 1} = 0,
  \label{eq:bcTy}
\end{align}
where the temperature at the bottom ($y = 0$) is higher than that at the top ($y = 1$).
The stream function $\psi(x, y, t)$ satisfies
the periodic boundary condition on $x$ and the Dirichlet zero boundary condition on $y$, i.e.,
\begin{align}
  & \psi(x + 2, y, t) = \psi(x, y, t),
  \label{eq:bcPx} \\
  & \Bigl. \psi(x, y, t) \Bigr|_{y = 0} = \Bigl. \psi(x, y, t) \Bigr|_{y = 1} = 0.
  \label{eq:bcPy}
\end{align}
Owing to the homogeneity of Eqs.~(\ref{eq:T})(\ref{eq:P_T})
and the periodic boundary condition on $x$, given in Eqs.~(\ref{eq:bcTx})(\ref{eq:bcPx}),
this system possesses continuous spatial translational symmetry with respect to $x$.
We also note that no conserved quantity exists in this system.

\subsection{Convective components of the temperature field}

To simplify the boundary conditions in Eq.~(\ref{eq:bcTy}),
we consider the convective component $X(x, y, t)$ of the temperature field $T(x, y, t)$ as follows:
\begin{align}
  T(x, y, t) = (1 - y) + X(x, y, t).
  \label{eq:T-X}
\end{align}
Substituting Eq.~(\ref{eq:T-X}) into Eq.~(\ref{eq:T}) and Eq.~(\ref{eq:P_T}),
we derive the following equations:
\begin{align}
  \frac{\partial}{\partial t} X(x, y, t)
  = \nabla^2 X + J(\psi, X) - \frac{\partial \psi}{\partial x},
  \label{eq:X}
\end{align}
and
\begin{align}
  \nabla^2 \psi(x, y, t) = -{\rm Ra} \frac{\partial X}{\partial x}.
  \label{eq:P_X}
\end{align}
Applying Eq.~(\ref{eq:T-X}) to Eqs.~(\ref{eq:bcTx})(\ref{eq:bcTy}),
we obtain the following boundary conditions for the convective component $X(x, y, t)$:
\begin{align}
  & X(x + 2, y, t) = X(x, y, t),
  \label{eq:bcXx} \\
  & \Bigl. X(x, y, t) \Bigr|_{y = 0} = \Bigl. X(x, y, t) \Bigr|_{y = 1} = 0.
  \label{eq:bcXy}
\end{align}
That is, the convective component $X(x, y, t)$ satisfies
the periodic boundary condition on $x$ and the Dirichlet zero boundary condition on $y$.

In the derivation below,
it should be noted that Eq.~(\ref{eq:P_X}) can also be written in the following form:
\begin{align}
  \psi(x, y, t)
  = \int_0^2 dx' \int_0^1 dy' \, G(x, y, x', y') \frac{\partial}{\partial x'} X( x', y', t),
  \label{eq:P}
\end{align}
where the Green's function $G(x, y, x', y')$ is the solution to
\begin{align}
  \nabla^2 G(x, y, x', y') = -{\rm Ra} \, \delta(x - x') \, \delta(y - y'),
  \label{eq:G}
\end{align}
under the periodic boundary condition on $x$ and the Dirichlet zero boundary condition on $y$.
From the translational symmetry with respect to $x$,
the Green's function $G(x, y, x', y')$ possesses the following property:
$G(x, y, x', y') = G(x - x', y, 0, y')$.
In the following two subsections,
we analyze the dynamical equation~(\ref{eq:X}) with Eq.~(\ref{eq:P_X}) or Eq.~(\ref{eq:P})
under the boundary conditions given by
Eqs.~(\ref{eq:bcXx})(\ref{eq:bcXy}) and Eqs.~(\ref{eq:bcPx})(\ref{eq:bcPy}).

\subsection{Limit-torus solution and its Floquet-type system}

\begin{figure*}
  \begin{center}
    \includegraphics[width=1.0\hsize,clip]{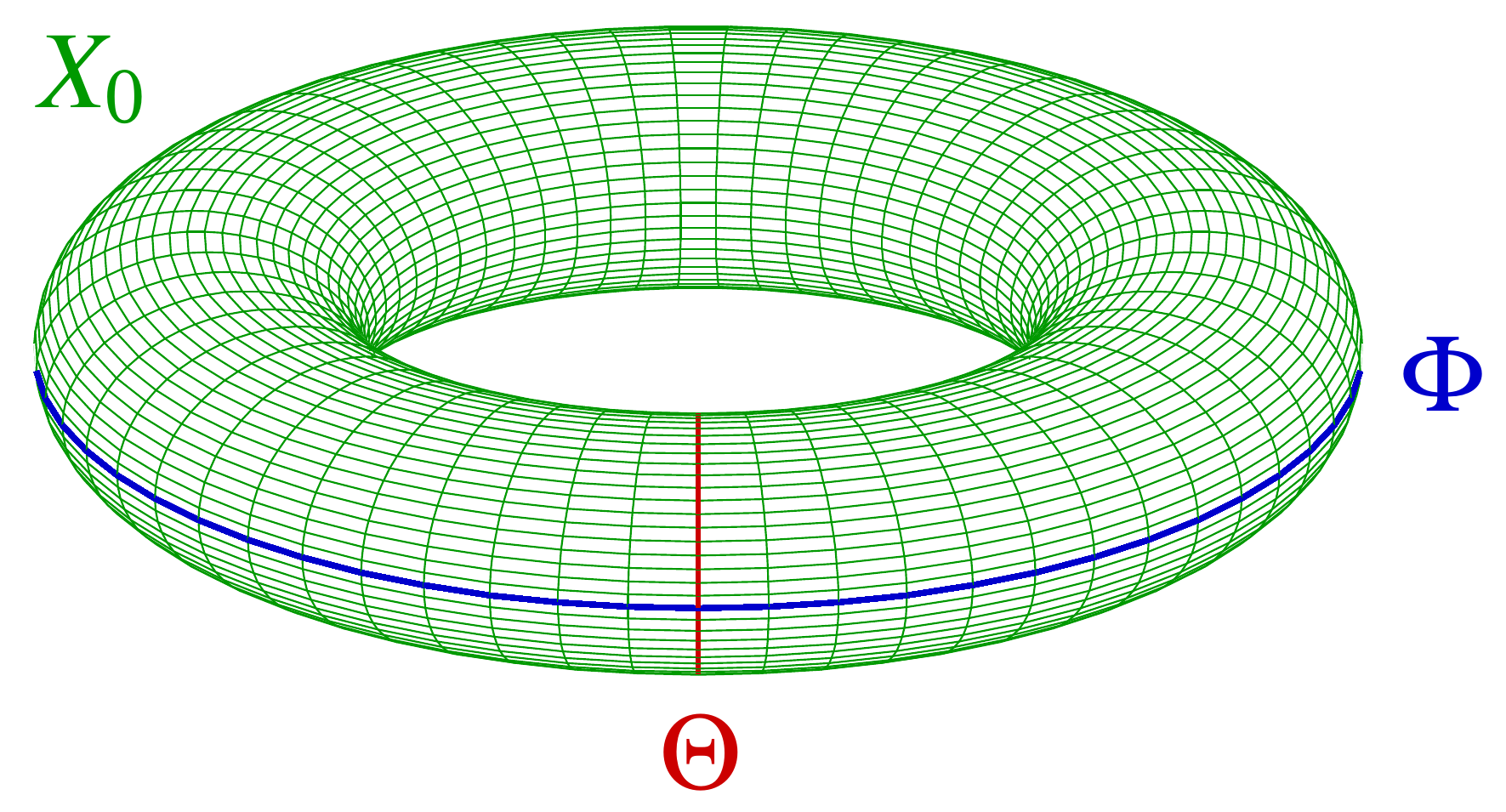}
  \end{center}
  \caption{(Color online)
    Schematic diagram of the limit-torus solution $X_0(x - \Phi, y, \Theta)$.
    The spatial phase $\Phi$ and temporal phase $\Theta$ are represented by
    the toroidal coordinate and poloidal coordinate, respectively.
  }
  \label{fig:2}
\end{figure*}

\begin{figure*}
  \begin{center}
    \includegraphics[width=1.0\hsize,clip]{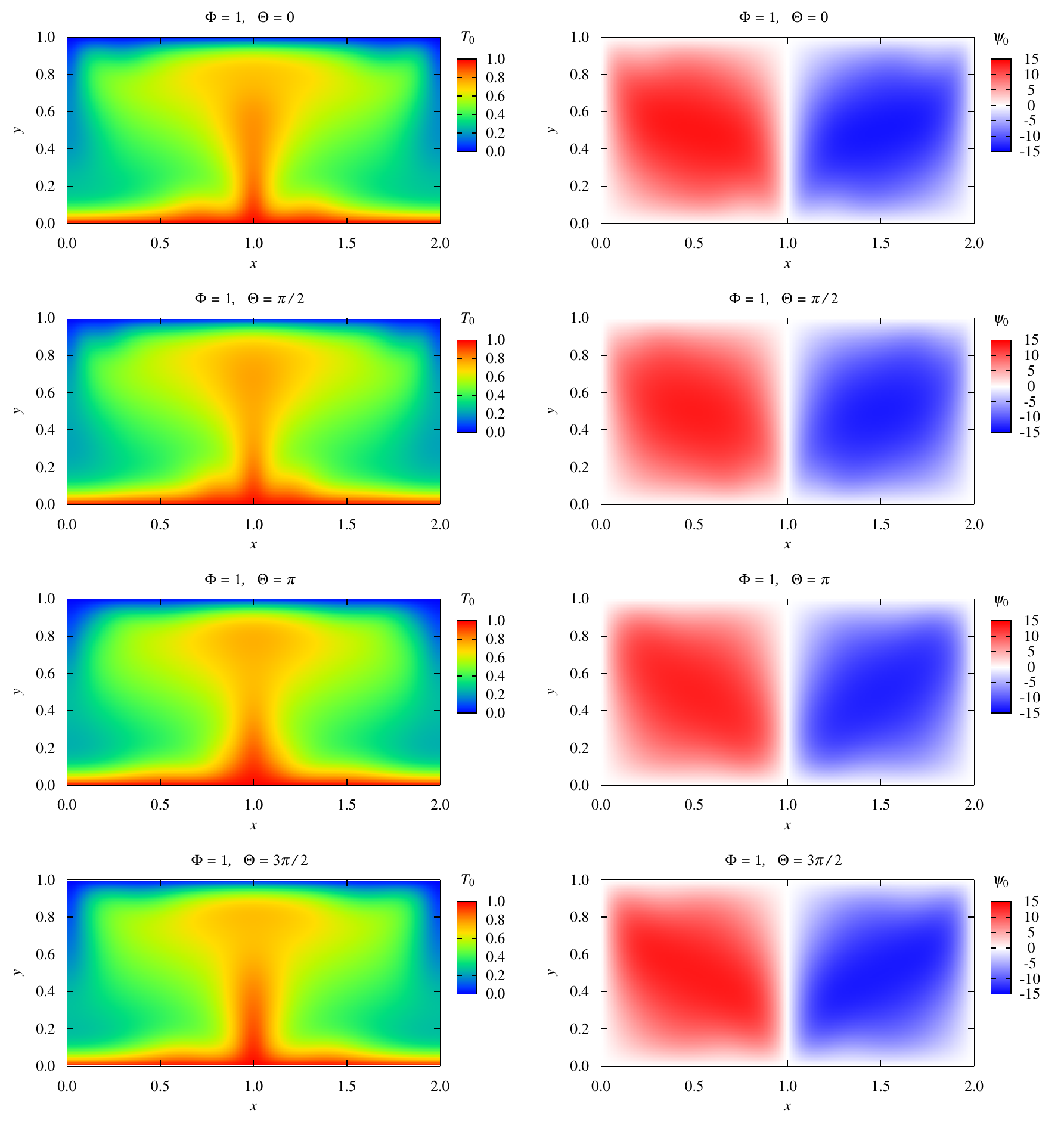}
  \end{center}
  \caption{(Color online)
    Snapshots of both $T_0(x - \Phi, y, \Theta)$
    and $\psi_0(x - \Phi, y, \Theta)$ with $\Phi = 1$
    for $\Theta = 0$, $\pi/2$, $\pi$, and $3\pi/2$.
  }
  \label{fig:5}
\end{figure*}

In general, a stable limit-torus solution to Eq.~(\ref{eq:X}),
which represents oscillatory convection in the cylindrical Hele-Shaw cell,
can be described by
(e.g., see Fig.~\ref{fig:4} in Sec.~\ref{sec:3})
\begin{align}
  X(x, y, t) = X_0\bigl( x - \Phi(t), y, \Theta(t) \bigr), \qquad
  \dot{\Phi}(t) = c, \qquad
  \dot{\Theta}(t) = \omega.
  \label{eq:X_X0}
\end{align}
The spatial phase and traveling velocity are denoted by $\Phi$ and $c$, respectively;
the temporal phase and oscillation frequency are denoted by $\Theta$ and $\omega$, respectively.
The spatial and temporal phases indicate the ``position'' and ``oscillation'' of the convection, respectively.
Figure~\ref{fig:2} shows a schematic diagram of the limit-torus solution $X_0(x - \Phi, y, \Theta)$.
The limit-torus solution $X_0(x - \Phi, y, \Theta)$ satisfies
the $2$-periodicity with respect to $\Phi$
and the $2\pi$-periodicity with respect to $\Theta$, i.e.,
\begin{align}
  X_0(x - \Phi + 2, y, \Theta) &= X_0(x - \Phi, y, \Theta),
  \\
  X_0(x - \Phi, y, \Theta + 2\pi) &= X_0(x - \Phi, y, \Theta).
\end{align}
Substituting Eq.~(\ref{eq:X_X0}) into Eqs.~(\ref{eq:X})(\ref{eq:P_X}),
we find that the limit-torus solution $X_0(x - \Phi, y, \Theta)$ satisfies the following equation:
\begin{align}
  \left[ - c \frac{\partial}{\partial x} + \omega \frac{\partial}{\partial \Theta} \right]
  X_0(x - \Phi, y, \Theta)
  = \nabla^2 X_0 + J(\psi_0, X_0) - \frac{\partial \psi_0}{\partial x},
  \label{eq:X0}
\end{align}
where the stream function $\psi_0(x - \Phi, y, \Theta)$ is determined by
(e.g., see Fig.~\ref{fig:5} in Sec.~\ref{sec:3})
\begin{align}
  \nabla^2 \psi_0(x - \Phi, y, \Theta) = -{\rm Ra} \frac{\partial X_0}{\partial x}.
  \label{eq:P0}
\end{align}
From Eq.~(\ref{eq:T-X}), the corresponding temperature field $T_0(x, y, \Theta)$ is given by
(e.g., see Fig.~\ref{fig:5} in Sec.~\ref{sec:3})
\begin{align}
  T_0(x - \Phi, y, \Theta) = (1 - y) + X_0(x - \Phi, y, \Theta).
  \label{eq:T0}
\end{align}

Let $u(x - \Phi, y, \Theta, t)$ represent a small disturbance
to the limit-torus solution $X_0(x - \Phi, y, \Theta)$,
and consider a slightly perturbed solution
\begin{align}
  X(x, y, t) = X_0\bigl( x - \Phi(t), y, \Theta(t) \bigr) + u\bigl( x - \Phi(t), y, \Theta(t), t \bigr).
\end{align}
Equation~(\ref{eq:X}) is then linearized with respect to $u(x - \Phi, y, \Theta, t)$ as follows:
\begin{align}
  \frac{\partial}{\partial t} u(x - \Phi, y, \Theta, t)
  = {\cal L}(x - \Phi, y, \Theta) u(x - \Phi, y, \Theta, t).
  \label{eq:linear}
\end{align}
Here, the linear operator ${\cal L}(x - \Phi, y, \Theta)$ is explicitly given by
\begin{align}
  {\cal L}(x - \Phi, y, \Theta) u(x - \Phi, y, \Theta)
  = \left[ L(x - \Phi, y, \Theta) + c \frac{\partial}{\partial x} - \omega \frac{\partial}{\partial \Theta} \right]
  u(x - \Phi, y, \Theta),
  \label{eq:calL}
\end{align}
where
\begin{align}
  L(x - \Phi, y, \Theta) u(x - \Phi, y, \Theta)
  = \nabla^2 u + J(\psi_0, u) + J(\psi_u, X_0) - \frac{\partial \psi_u}{\partial x}.
  \label{eq:L}
\end{align}
Similarly to the limit-torus solution $X_0(x - \Phi, y, \Theta)$,
the function $u(x - \Phi, y, \Theta)$ satisfies
the periodic boundary condition on $x$,
the Dirichlet zero boundary condition on $y$,
and the $2\pi$-periodicity with respect to $\Theta$.
In Eq.~(\ref{eq:L}), the function $\psi_u(x - \Phi, y, \Theta)$ is the solution to
\begin{align}
  \nabla^2 \psi_u(x - \Phi, y, \Theta) = -{\rm Ra} \frac{\partial u}{\partial x},
  \label{eq:P_u}
\end{align}
under the periodic boundary condition on $x$ and the Dirichlet zero boundary condition on $y$.
Note that the linear operator ${\cal L}(x - \Phi, y, \Theta)$
is periodic with respect to both $\Phi$ and $\Theta$.
Therefore, Eq.~(\ref{eq:linear}) is a Floquet-type system with two zero-eigenvalues;
one is associated with spatial translational symmetry breaking
and the other is associated with temporal translational symmetry breaking.

Defining the inner product of two functions as
\begin{align}
  \Pd{ u^\ast(x - \Phi, y, \Theta), \, u(x - \Phi, y, \Theta) }
  = \frac{1}{2\pi} \int_0^{2\pi} d\Theta \int_0^2 dx \int_0^1 dy \,
  u^\ast(x - \Phi, y, \Theta) u(x - \Phi, y, \Theta),
  \label{eq:inner}
\end{align}
we introduce the adjoint operator of the linear operator ${\cal L}(x - \Phi, y, \Theta)$ by
\begin{align}
  \Pd{ u^\ast(x - \Phi, y, \Theta), \, {\cal L}(x - \Phi, y, \Theta) u(x - \Phi, y, \Theta) }
  = \Pd{ {\cal L}^\ast(x - \Phi, y, \Theta) u^\ast(x - \Phi, y, \Theta), \, u(x - \Phi, y, \Theta) }.
  \label{eq:operator}
\end{align}
By partial integration, the adjoint operator ${\cal L}^\ast(x - \Phi, y, \Theta)$ is explicitly given by
\begin{align}
  {\cal L}^\ast(x - \Phi, y, \Theta) u^\ast(x - \Phi, y, \Theta)
  = \left[ L^\ast(x - \Phi, y, \Theta) - c \frac{\partial}{\partial x} + \omega \frac{\partial}{\partial \Theta} \right]
  u^\ast(x - \Phi, y, \Theta),
  \label{eq:calLast}
\end{align}
where
\begin{align}
  L^\ast(x - \Phi, y, \Theta) u^\ast(x - \Phi, y, \Theta)
  = \nabla^2 u^\ast
  + \frac{\partial}{\partial x} \left[ u^\ast \frac{\partial \psi_0}{\partial y} \right]
  - \frac{\partial}{\partial y} \left[ u^\ast \frac{\partial \psi_0}{\partial x} \right]
  + \frac{\partial}{\partial x} \Bigl[ \psi_{u,x}^\ast - \psi_{u,y}^\ast \Bigr].
  \label{eq:Last}
\end{align}
Similarly to $u(x - \Phi, y, \Theta)$,
the function $u^\ast(x - \Phi, y, \Theta)$ also satisfies
the periodic boundary condition on $x$,
the Dirichlet zero boundary condition on $y$,
and the $2\pi$-periodicity with respect to $\Theta$.
In Eq.~(\ref{eq:Last}), the two functions, i.e.,
$\psi_{u,x}^\ast(x, y, \Theta)$ and $\psi_{u,y}^\ast(x, y, \Theta)$,
are the solutions to
\begin{align}
  \nabla^2 \psi_{u,x}^\ast(x - \Phi, y, \Theta)
  &= -{\rm Ra} \frac{\partial}{\partial x}
  \left[ u^\ast \left( \frac{\partial X_0}{\partial y} - 1 \right) \right],
  \label{eq:Past_uast_x} \\
  \nabla^2 \psi_{u,y}^\ast(x - \Phi, y, \Theta)
  &= -{\rm Ra} \frac{\partial}{\partial y}
  \left[ u^\ast \frac{\partial X_0}{\partial x} \right],
  \label{eq:Past_uast_y}
\end{align}
under the periodic boundary condition on $x$ and the Dirichlet zero boundary condition on $y$, respectively.
Details of the derivation of the adjoint operator ${\cal L}^\ast(x - \Phi, y, \Theta)$
are given in App.~\ref{sec:A}.

\subsection{Floquet zero eigenfunctions}

In the calculation below,
we utilize the Floquet eigenfunctions associated with the two zero-eigenvalues, i.e.,
\begin{align}
  {\cal L}(x - \Phi, y, \Theta) U_{\rm s}(x - \Phi, y, \Theta)
  = \left[ L(x - \Phi, y, \Theta) + c \frac{\partial}{\partial x} - \omega \frac{\partial}{\partial \Theta} \right]
  U_{\rm s}(x - \Phi, y, \Theta)
  &= 0,
  \\
  {\cal L}(x - \Phi, y, \Theta) U_{\rm t}(x - \Phi, y, \Theta)
  = \left[ L(x - \Phi, y, \Theta) + c \frac{\partial}{\partial x} - \omega \frac{\partial}{\partial \Theta} \right]
  U_{\rm t}(x - \Phi, y, \Theta)
  &= 0,
  \\
  {\cal L}^\ast(x - \Phi, y, \Theta) U_{\rm s}^\ast(x - \Phi, y, \Theta)
  = \left[ L^\ast(x - \Phi, y, \Theta) - c \frac{\partial}{\partial x} + \omega \frac{\partial}{\partial \Theta} \right]
  U_{\rm s}^\ast(x - \Phi, y, \Theta)
  &= 0,
  \label{eq:Usast} \\
  {\cal L}^\ast(x - \Phi, y, \Theta) U_{\rm t}^\ast(x - \Phi, y, \Theta)
  = \left[ L^\ast(x - \Phi, y, \Theta) - c \frac{\partial}{\partial x} + \omega \frac{\partial}{\partial \Theta} \right]
  U_{\rm t}^\ast(x - \Phi, y, \Theta)
  &= 0.
  \label{eq:Utast}
\end{align}
Note that the two right zero eigenfunctions, i.e.,
$U_{\rm s}(x - \Phi, y, \Theta)$ for the spatial phase $\Phi$
and
$U_{\rm t}(x - \Phi, y, \Theta)$ for the temporal phase $\Theta$,
can be chosen as
(e.g., see Fig.~\ref{fig:7} in Sec.~\ref{sec:3})
\begin{align}
  U_{\rm s}(x - \Phi, y, \Theta)
  &= \frac{\partial}{\partial x} X_0(x - \Phi, y, \Theta),
  \label{eq:Us} \\
  U_{\rm t}(x - \Phi, y, \Theta)
  &= \frac{\partial}{\partial \Theta} X_0(x - \Phi, y, \Theta),
  \label{eq:Ut}
\end{align}
which are confirmed by differentiating Eq.~(\ref{eq:X0}) with respect to $\Phi$ and $\Theta$, respectively.
For the inner product~(\ref{eq:inner}) with the two right zero eigenfunctions~(\ref{eq:Us})(\ref{eq:Ut}),
the corresponding two left zero eigenfunctions, i.e.,
$U_{\rm s}^\ast(x - \Phi, y, \Theta)$ and $U_{\rm t}^\ast(x - \Phi, y, \Theta)$,
are orthonormalized as
\begin{align}
  \Pd{ U_p^\ast(x - \Phi, y, \Theta), \, U_q(x - \Phi, y, \Theta) }
  = \frac{1}{2\pi} \int_0^{2\pi} d\Theta \int_0^2 dx \int_0^1 dy \,
  U_p^\ast(x - \Phi, y, \Theta) U_q(x - \Phi, y, \Theta)
  = \delta_{pq},
\end{align}
for $p, q = {\rm s}, {\rm t}$.
Here, we note that the following equation holds
(see also Refs.~\cite{ref:kawamura11,ref:kawamura13,ref:hoppensteadt97}):
\begin{align}
  & \frac{\partial}{\partial \Theta}
  \left[ \int_0^2 dx \int_0^1 dy \, U_p^\ast(x - \Phi, y, \Theta) U_q(x - \Phi, y, \Theta) \right]
  \nonumber \\
  &= \int_0^2 dx \int_0^1 dy \, \left[
    U_p^\ast(x - \Phi, y, \Theta) \frac{\partial}{\partial \Theta} U_q(x - \Phi, y, \Theta)
    + U_q(x - \Phi, y, \Theta) \frac{\partial}{\partial \Theta} U_p^\ast(x - \Phi, y, \Theta) \right]
  \nonumber \\
  &= \frac{1}{\omega} \int_0^2 dx \int_0^1 dy \, \biggl[
    U_p^\ast(x - \Phi, y, \Theta) \left\{ L(x - \Phi, y, \Theta) + c \frac{\partial}{\partial x} \right\} U_q(x - \Phi, y, \Theta)
    \nonumber \\
    & \qquad\qquad\qquad\qquad\quad
    - U_q(x - \Phi, y, \Theta) \left\{ L^\ast(x - \Phi, y, \Theta) - c \frac{\partial}{\partial x} \right\} U_p^\ast(x - \Phi, y, \Theta)
    \biggr]
  \nonumber \\
  &= 0.
\end{align}
Therefore, the following orthonormalization condition is independently satisfied for each $\Theta$:
\begin{align}
  \int_0^2 dx \int_0^1 dy \, U_p^\ast(x - \Phi, y, \Theta) U_q(x - \Phi, y, \Theta) = \delta_{pq}.
  \label{eq:orthonormal}
\end{align}

Here, we describe a numerical method for obtaining the left zero eigenfunctions.
From Eqs.~(\ref{eq:Usast})(\ref{eq:Utast}), the left zero eigenfunctions,
$U_{\rm s}^\ast(x - \Phi, y, \Theta)$ and $U_{\rm t}^\ast(x - \Phi, y, \Theta)$, satisfy
\begin{align}
  \omega \frac{\partial}{\partial \Theta} U_p^\ast(x - \Phi, y, \Theta)
  = - \left[ L^\ast(x - \Phi, y, \Theta) - c \frac{\partial}{\partial x} \right] U_p^\ast(x - \Phi, y, \Theta),
\end{align}
for $p = {\rm s}, {\rm t}$,
which can be transformed into
\begin{align}
  \frac{\partial}{\partial s} U_p^\ast(x - \Phi, y, -\omega s)
  = \left[ L^\ast(x - \Phi, y, -\omega s) - c \frac{\partial}{\partial x} \right] U_p^\ast(x - \Phi, y, -\omega s),
  \label{eq:adjoint}
\end{align}
by substituting $\Theta = -\omega s$.
A relaxation method using Eq.~(\ref{eq:adjoint}),
which can also be called the adjoint method
(
see Refs.~\cite{ref:hoppensteadt97,ref:izhikevich07,ref:ermentrout10,ref:ermentrout96,ref:brown04}
for limit-cycle solutions to ordinary differential equations
and Refs.~\cite{ref:kawamura11,ref:kawamura13,ref:nakao12}
for limit-cycle solutions to partial differential equations
),
is convenient to obtain the left zero eigenfunctions for the limit-torus solution.
In the following two subsections,
we derive a set of phase equations for oscillatory convection in the cylindrical Hele-Shaw cell
using the limit-torus solution and its Floquet zero eigenfunctions.

\subsection{Oscillatory convection with weak perturbations} \label{subsec:2E}

In this subsection,
we consider oscillatory cylindrical-Hele-Shaw convection
with a weak perturbation to the temperature field $T(x, y, t)$
described by the following equation:
\begin{align}
  \frac{\partial}{\partial t} T(x, y, t) = \nabla^2 T + J(\psi, T) + \epsilon p(x, y, t).
  \label{eq:Tp}
\end{align}
The weak perturbation is denoted by $\epsilon p(x, y, t)$.
Substituting Eq.~(\ref{eq:T-X}) into Eq.~(\ref{eq:Tp}),
we obtain the following equation for the convective component $X(x, y, t)$:
\begin{align}
  \frac{\partial}{\partial t} X(x, y, t)
  = \nabla^2 X + J(\psi, X) - \frac{\partial \psi}{\partial x} + \epsilon p(x, y, t).
  \label{eq:Xp}
\end{align}
Using the idea of phase reduction~\cite{ref:kuramoto84},
we derive a set of phase equations from the perturbed equation~(\ref{eq:Xp}).
That is, using the left zero eigenfunctions,
i.e., $U_{\rm s}^\ast(x - \Phi, y, \Theta)$ and $U_{\rm t}^\ast(x - \Phi, y, \Theta)$,
we project the dynamics of the perturbed equation~(\ref{eq:Xp})
onto the unperturbed limit-torus solution with respect to the spatial and temporal phases as follows:
\begin{align}
  -\dot{\Phi}(t)
  &= \int_0^2 dx \int_0^1 dy \, U_{\rm s}^\ast(x - \Phi, y, \Theta)
  \left[ \frac{\partial}{\partial t} X(x, y, t) \right]
  \nonumber \\
  &= \int_0^2 dx \int_0^1 dy \, U_{\rm s}^\ast(x - \Phi, y, \Theta) 
  \left[ \nabla^2 X + J(\psi, X) - \frac{\partial \psi}{\partial x} + \epsilon p(x, y, t) \right]
  \nonumber \\
  &\simeq \int_0^2 dx \int_0^1 dy \, U_{\rm s}^\ast(x - \Phi, y, \Theta)
  \left[ \nabla^2 X_0 + J(\psi_0, X_0) - \frac{\partial \psi_0}{\partial x} + \epsilon p(x, y, t) \right]
  \nonumber \\
  &= \int_0^2 dx \int_0^1 dy \, U_{\rm s}^\ast(x - \Phi, y, \Theta)
  \left[ - c \frac{\partial X_0}{\partial x} + \omega \frac{\partial X_0}{\partial \Theta} + \epsilon p(x, y, t) \right]
  \nonumber \\
  &= \int_0^2 dx \int_0^1 dy \, U_{\rm s}^\ast(x - \Phi, y, \Theta)
  \biggl[ - c U_{\rm s}(x - \Phi, y , \Theta) + \omega U_{\rm t}(x - \Phi, y, \Theta) + \epsilon p(x, y, t) \biggr]
  \nonumber \\
  &= -c + \epsilon \int_0^2 dx \int_0^1 dy \, U_{\rm s}^\ast(x - \Phi, y, \Theta) p(x, y, t),
  \label{eq:project_Phi}
\end{align}
and
\begin{align}
  \dot{\Theta}(t)
  &= \int_0^2 dx \int_0^1 dy \, U_{\rm t}^\ast(x - \Phi, y, \Theta)
  \left[ \frac{\partial}{\partial t} X(x, y, t) \right]
  \nonumber \\
  &= \int_0^2 dx \int_0^1 dy \, U_{\rm t}^\ast(x - \Phi, y, \Theta) 
  \left[ \nabla^2 X + J(\psi, X) - \frac{\partial \psi}{\partial x} + \epsilon p(x, y, t) \right]
  \nonumber \\
  &\simeq \int_0^2 dx \int_0^1 dy \, U_{\rm t}^\ast(x - \Phi, y, \Theta)
  \left[ \nabla^2 X_0 + J(\psi_0, X_0) - \frac{\partial \psi_0}{\partial x} + \epsilon p(x, y, t) \right]
  \nonumber \\
  &= \int_0^2 dx \int_0^1 dy \, U_{\rm t}^\ast(x - \Phi, y, \Theta)
  \left[ - c \frac{\partial X_0}{\partial x} + \omega \frac{\partial X_0}{\partial \Theta} + \epsilon p(x, y, t) \right]
  \nonumber \\
  &= \int_0^2 dx \int_0^1 dy \, U_{\rm t}^\ast(x - \Phi, y, \Theta)
  \biggl[ - c U_{\rm s}(x - \Phi, y , \Theta) + \omega U_{\rm t}(x - \Phi, y, \Theta) + \epsilon p(x, y, t) \biggr]
  \nonumber \\
  &= \omega + \epsilon \int_0^2 dx \int_0^1 dy \, U_{\rm t}^\ast(x - \Phi, y, \Theta) p(x, y, t),
  \label{eq:project_Theta}
\end{align}
where we approximated $X(x, y, t)$ by the unperturbed limit-torus solution $X_0(x - \Phi, y, \Theta)$
in Eqs.~(\ref{eq:project_Phi})(\ref{eq:project_Theta}).
Therefore, the two phase equations describing the oscillatory cylindrical-Hele-Shaw convection with weak perturbation
are approximately obtained in the following forms:
\begin{align}
  \dot{\Phi}(t) &= c + \epsilon \int_0^2 dx \int_0^1 dy \, Z_{\rm s}(x - \Phi, y, \Theta) p(x, y, t),
  \label{eq:Phi_p} \\
  \dot{\Theta}(t) &= \omega + \epsilon \int_0^2 dx \int_0^1 dy \, Z_{\rm t}(x - \Phi, y, \Theta) p(x, y, t),
  \label{eq:Theta_p}
\end{align}
where the {\it phase sensitivity functions} for the spatial and temporal phases are defined as
(e.g., see Fig.~\ref{fig:8} in Sec.~\ref{sec:3})
\begin{align}
  Z_{\rm s}(x - \Phi, y, \Theta) &= -U_{\rm s}^\ast(x - \Phi, y, \Theta),
  \label{eq:Zs} \\
  Z_{\rm t}(x - \Phi, y, \Theta) &=  U_{\rm t}^\ast(x - \Phi, y, \Theta).
  \label{eq:Zt}
\end{align}
The phase equations~(\ref{eq:Phi_p})(\ref{eq:Theta_p}) are the main results of this paper.
These equations can also be considered as an extension of that
describing oscillatory convection in the ordinary Hele-Shaw cell~\cite{ref:kawamura13}.
As found from Eqs.~(\ref{eq:Phi_p})(\ref{eq:Theta_p}),
the spatial and temporal phases are coupled;
therefore, nontrivial spatiotemporal phase dynamics are revealed.

Furthermore, we consider the case of the perturbation described by a product of two functions as follows:
\begin{align}
  p(x, y, t) = a(x, y) q(t).
  \label{eq:p_aq}
\end{align}
That is, the space-dependence and time-dependence of the perturbation are separated.
In this case, the phase equations~(\ref{eq:Phi_p})(\ref{eq:Theta_p}) are written in the following forms:
\begin{align}
  \dot{\Phi}(t) &= c + \epsilon \zeta_{\rm s}(\Phi, \Theta) q(t),
  \label{eq:Phi_q} \\
  \dot{\Theta}(t) &= \omega + \epsilon \zeta_{\rm t}(\Phi, \Theta) q(t),
  \label{eq:Theta_q}
\end{align}
where the {\it effective phase sensitivity functions} for the spatial and temporal phases are given by
(e.g., see Fig.~\ref{fig:10} in Sec.~\ref{sec:3})
\begin{align}
  \zeta_{\rm s}(\Phi, \Theta) &= \int_0^2 dx \int_0^1 dy \, Z_{\rm s}(x - \Phi, y, \Theta) a(x, y),
  \label{eq:zetas} \\
  \zeta_{\rm t}(\Phi, \Theta) &= \int_0^2 dx \int_0^1 dy \, Z_{\rm t}(x - \Phi, y, \Theta) a(x, y).
  \label{eq:zetat}
\end{align}
We note that the forms of Eqs.~(\ref{eq:Phi_q})(\ref{eq:Theta_q}) are essentially the same as
those of the phase equations which are derived for a perturbed limit-torus oscillator
described by a finite-dimensional dynamical system
(see Refs.~\cite{ref:izhikevich99,ref:demir10}).

Finally, it should be noted that
we can also consider oscillatory cylindrical-Hele-Shaw convection with weak boundary forcing
as mentioned in App.~\ref{sec:B}.

\subsection{Weakly coupled systems of oscillatory convection} \label{subsec:2F}

In this subsection,
we consider weakly coupled systems of oscillatory cylindrical-Hele-Shaw convection
described by the following equation:
\begin{align}
  \frac{\partial}{\partial t} T_\sigma(x, y, t)
  = \nabla^2 T_\sigma + J(\psi_\sigma, T_\sigma) + \epsilon \bigl( T_\tau - T_\sigma \bigr),
  \label{eq:Tsigma}
\end{align}
for $(\sigma, \tau) = (1, 2)$ or $(2, 1)$.
Two identical systems of oscillatory cylindrical-Hele-Shaw convection are mutually coupled
through corresponding temperatures at each spatial point~\footnote{
  As in Ref.~\cite{ref:kawamura13},
  the phase description method developed in this paper is also applicable to any coupling form,
  e.g., asymmetric, nonlinear, spatially nonlocal, or spatially partial coupling,
  as long as the coupling intensity is sufficiently weak.
},
where the coupling parameter is denoted by $\epsilon$.
Substituting Eq.~(\ref{eq:T-X}) into Eq.~(\ref{eq:Tsigma}) for each $\sigma$,
we obtain the following equation for the convective component $X_\sigma(x, y, t)$:
\begin{align}
  \frac{\partial}{\partial t} X_\sigma(x, y, t)
  = \nabla^2 X_\sigma + J(\psi_\sigma, X_\sigma) - \frac{\partial \psi_\sigma}{\partial x}
  + \epsilon \bigl( X_\tau - X_\sigma \bigr).
  \label{eq:Xsigma}
\end{align}
As in Eq.~(\ref{eq:P_X}), the stream function $\psi_\sigma(x, y, t)$ of each system is determined by
\begin{align}
  \nabla^2 \psi_\sigma(x, y, t) = -{\rm Ra} \frac{\partial X_\sigma}{\partial x}.
\end{align}

Here, we assume that unperturbed oscillatory cylindrical-Hele-Shaw convection is a stable limit-torus solution
and that the coupling between the systems of oscillatory convection is sufficiently weak.
Under this assumption, as in the preceding subsection,
we obtain a set of phase equations from Eq.~(\ref{eq:Xsigma}) as follows:
\begin{align}
  \dot{\Phi}_\sigma(t)
  &= c + \epsilon \tilde{\Gamma}_{\rm s}\left( \Phi_\sigma - \Phi_\tau, \Theta_\sigma, \Theta_\tau \right),
  \label{eq:Phi_tildeGamma} \\
  \dot{\Theta}_\sigma(t)
  &= \omega + \epsilon \tilde{\Gamma}_{\rm t}\left( \Phi_\sigma - \Phi_\tau, \Theta_\sigma, \Theta_\tau \right),
  \label{eq:Theta_tildeGamma}
\end{align}
where
\begin{align}
  \tilde{\Gamma}_{\rm s}\left( \Phi_\sigma - \Phi_\tau, \Theta_\sigma, \Theta_\tau \right)
  &= \int_0^2 dx \int_0^1 dy \, Z_{\rm s}(x - \Phi_\sigma, y, \Theta_\sigma)
  \Bigl[ X_0(x - \Phi_\tau, y, \Theta_\tau) - X_0(x - \Phi_\sigma, y, \Theta_\sigma) \Bigr],
  \\
  \tilde{\Gamma}_{\rm t}\left( \Phi_\sigma - \Phi_\tau, \Theta_\sigma, \Theta_\tau \right)
  &= \int_0^2 dx \int_0^1 dy \, Z_{\rm t}(x - \Phi_\sigma, y, \Theta_\sigma)
  \Bigl[ X_0(x - \Phi_\tau, y, \Theta_\tau) - X_0(x - \Phi_\sigma, y, \Theta_\sigma) \Bigr].
\end{align}
These two functions, i.e.,
$\tilde{\Gamma}_{\rm s}(\Phi_\sigma - \Phi_\tau, \Theta_\sigma, \Theta_\tau)$
and
$\tilde{\Gamma}_{\rm t}(\Phi_\sigma - \Phi_\tau, \Theta_\sigma, \Theta_\tau)$,
depend on the spatial phase difference between the systems
and the temporal phases of both systems.

Introducing the slow phase variables as 
\begin{align}
  \Phi_\sigma(t) &= c t + \phi_\sigma(t),
  \\
  \Theta_\sigma(t) &= \omega t + \theta_\sigma(t),
\end{align}
we rewrite Eqs.~(\ref{eq:Phi_tildeGamma})(\ref{eq:Theta_tildeGamma}) as
\begin{align}
  \dot{\phi}_\sigma(t)
  &= \epsilon \tilde{\Gamma}_{\rm s}\left( \phi_\sigma - \phi_\tau, \omega t + \theta_\sigma, \omega t + \theta_\tau \right),
  \label{eq:phi_tildeGamma} \\
  \dot{\theta}_\sigma(t)
  &= \epsilon \tilde{\Gamma}_{\rm t}\left( \phi_\sigma - \phi_\tau, \omega t + \theta_\sigma, \omega t + \theta_\tau \right).
  \label{eq:theta_tildeGamma}
\end{align}
By applying the averaging method with respect to the temporal phases,
Eqs.~(\ref{eq:phi_tildeGamma})(\ref{eq:theta_tildeGamma}) are written in the following forms:
\begin{align}
  \dot{\phi}_\sigma(t)
  &= \epsilon \Gamma_{\rm s}\left( \phi_\sigma - \phi_\tau, \theta_\sigma - \theta_\tau \right),
  \\
  \dot{\theta}_\sigma(t)
  &= \epsilon \Gamma_{\rm t}\left( \phi_\sigma - \phi_\tau, \theta_\sigma - \theta_\tau \right),
\end{align}
where the {\it phase coupling functions} for the spatial and temporal phases are given by
(e.g., see Fig.~\ref{fig:12} in Sec.~\ref{sec:3})
\begin{align}
  \Gamma_{\rm s}\left( \phi_\sigma - \phi_\tau, \theta_\sigma - \theta_\tau \right)
  &= \frac{1}{2\pi} \int_0^{2\pi} d\lambda \,
  \tilde{\Gamma}_{\rm s}\left( \phi_\sigma - \phi_\tau, \lambda + \theta_\sigma, \lambda + \theta_\tau \right),
  \\
  \Gamma_{\rm t}\left( \phi_\sigma - \phi_\tau, \theta_\sigma - \theta_\tau \right)
  &= \frac{1}{2\pi} \int_0^{2\pi} d\lambda \,
  \tilde{\Gamma}_{\rm t}\left( \phi_\sigma - \phi_\tau, \lambda + \theta_\sigma, \lambda + \theta_\tau \right).
\end{align}
Therefore, we obtain the following phase equations:
\begin{align}
  \dot{\Phi}_\sigma(t)
  &= c + \epsilon \Gamma_{\rm s}\left( \Phi_\sigma - \Phi_\tau, \Theta_\sigma - \Theta_\tau \right),
  \label{eq:Phi_Gamma} \\
  \dot{\Theta}_\sigma(t)
  &= \omega + \epsilon \Gamma_{\rm t}\left( \Phi_\sigma - \Phi_\tau, \Theta_\sigma - \Theta_\tau \right),
  \label{eq:Theta_Gamma}
\end{align}
where the phase coupling functions are explicitly described as
\begin{align}
  & \Gamma_{\rm s}\left( \Phi_\sigma - \Phi_\tau, \Theta_\sigma - \Theta_\tau \right)
  \nonumber \\
  & = \frac{1}{2\pi} \int_0^{2\pi} d\lambda \int_0^2 dx \int_0^1 dy \, Z_{\rm s}(x - \Phi_\sigma, y, \lambda + \Theta_\sigma)
  \Bigl[ X_0(x - \Phi_\tau, y, \lambda + \Theta_\tau) - X_0(x - \Phi_\sigma, y, \lambda + \Theta_\sigma) \Bigr],
  \label{eq:Gamma_s}\\[3mm]
  & \Gamma_{\rm t}\left( \Phi_\sigma - \Phi_\tau, \Theta_\sigma - \Theta_\tau \right)
  \nonumber \\
  & = \frac{1}{2\pi} \int_0^{2\pi} d\lambda \int_0^2 dx \int_0^1 dy \, Z_{\rm t}(x - \Phi_\sigma, y, \lambda + \Theta_\sigma)
  \Bigl[ X_0(x - \Phi_\tau, y, \lambda + \Theta_\tau) - X_0(x - \Phi_\sigma, y, \lambda + \Theta_\sigma) \Bigr].
  \label{eq:Gamma_t}
\end{align}
The phase coupling functions for the spatial and temporal phases, i.e.,
$\Gamma_{\rm s}(\Phi_\sigma - \Phi_\tau, \Theta_\sigma - \Theta_\tau)$
and
$\Gamma_{\rm t}(\Phi_\sigma - \Phi_\tau, \Theta_\sigma - \Theta_\tau)$,
depend only on the spatial and temporal phase differences.

Let the spatial and temporal phase differences respectively be defined as
\begin{align}
  \varDelta\Phi(t) &= \Phi_1(t) - \Phi_2(t),
  \\
  \varDelta\Theta(t) &= \Theta_1(t) - \Theta_2(t).
\end{align}
From Eqs.~(\ref{eq:Phi_Gamma})(\ref{eq:Theta_Gamma}), we obtain the following equations by subtraction:
\begin{align}
  \frac{d}{dt} \varDelta\Phi(t)
  &= \epsilon \Gamma_{\rm s}^{\rm (a)}\left( \varDelta\Phi, \varDelta\Theta \right),
  \label{eq:DPhi} \\
  \frac{d}{dt} \varDelta\Theta(t)
  &= \epsilon \Gamma_{\rm t}^{\rm (a)}\left( \varDelta\Phi, \varDelta\Theta \right),
  \label{eq:DTheta}
\end{align}
where
\begin{align}
  \Gamma_{\rm s}^{\rm (a)}\left( \varDelta\Phi, \varDelta\Theta \right)
  &= \Gamma_{\rm s}\left( \varDelta\Phi, \varDelta\Theta \right)
  - \Gamma_{\rm s}\left( -\varDelta\Phi, -\varDelta\Theta \right),
  \\
  \Gamma_{\rm t}^{\rm (a)}\left( \varDelta\Phi, \varDelta\Theta \right)
  &= \Gamma_{\rm t}\left( \varDelta\Phi, \varDelta\Theta \right)
  - \Gamma_{\rm t}\left( -\varDelta\Phi, -\varDelta\Theta \right).
\end{align}
These two functions, i.e.,
$\Gamma_{\rm s}^{\rm (a)}(\varDelta\Phi, \varDelta\Theta)$
and
$\Gamma_{\rm t}^{\rm (a)}(\varDelta\Phi, \varDelta\Theta)$,
satisfy the following properties:
\begin{align}
  \Gamma_{\rm s}^{\rm (a)}\left( -\varDelta\Phi, -\varDelta\Theta \right)
  &= -\Gamma_{\rm s}^{\rm (a)}\left( \varDelta\Phi, \varDelta\Theta \right),
  \label{eq:Gsa_as} \\
  \Gamma_{\rm t}^{\rm (a)}\left( -\varDelta\Phi, -\varDelta\Theta \right)
  &= -\Gamma_{\rm t}^{\rm (a)}\left( \varDelta\Phi, \varDelta\Theta \right),
  \label{eq:Gta_as}
\end{align}
which represent the anti-symmetry with respect to the origin, i.e.,
$\varDelta\Phi = \varDelta\Theta = 0$.

Finally, we note that the forms of Eqs.~(\ref{eq:Phi_Gamma})(\ref{eq:Theta_Gamma}) are the same as
that of the phase equations which are derived from weakly coupled limit-torus oscillators
described by finite-dimensional dynamical systems
(see Refs.~\cite{ref:izhikevich99,ref:demir10}).
That is, a system of oscillatory convection with a spatially translational mode
can be reduced to a set of phase equations,
similarly to an ordinary limit-torus oscillator.

\section{Numerical analysis of oscillatory convection} \label{sec:3}

In this section,
we illustrate the theory developed in Sec.~\ref{sec:2}
using a numerical analysis of the oscillatory convection in the cylindrical Hele-Shaw cell.

\subsection{Spectral transformation}

For numerical simulations using the pseudospectral method performed in Sec.~\ref{subsec:3B},
we first describe a spectral transformation.
Considering the boundary conditions for $X(x, y, t)$, given in Eqs.~(\ref{eq:bcXx})(\ref{eq:bcXy}),
we introduce the following spectral decomposition:
\begin{align}
  X(x, y, t) = \sum_{j=-\infty}^{\infty} \sum_{k=1}^{\infty} H_{jk}(t) \exp(i \pi j x) \sin(\pi k y),
\end{align}
where the spectral transformation of $X(x, y, t)$ is given by
\begin{align}
  H_{jk}(t) = \int_0^2 dx \int_0^1 dy \, X(x, y, t) \exp(-i \pi j x) \sin(\pi k y).
\end{align}
Similarly, the limit-torus solution $X_0(x - \Phi, y, \Theta)$ introduced in Eq.~(\ref{eq:X_X0}) is decomposed as
\begin{align}
  X_0(x - \Phi, y, \Theta)
  = \sum_{j=-\infty}^{\infty} \sum_{k=1}^{\infty} H_{jk}(\Theta) \exp(-i \pi j \Phi) \exp(i \pi j x) \sin(\pi k y).
\end{align}
The corresponding spectral transformation of the limit-torus solution is described by
\begin{align}
  H_{jk}(\Theta) \exp(-i \pi j \Phi)
  = \int_0^2 dx \int_0^1 dy \, X_0(x - \Phi, y, \Theta) \exp(-i \pi j x) \sin(\pi k y),
  \label{eq:Hjk}
\end{align}
where we assign the origin of the spatial phase, i.e., $\Phi = 0$,
to the spatial pattern $X_0(x, y, \Theta)$ that satisfies the following property:
\begin{align}
  H_{-1,1}(\Theta) = \int_0^2 dx \int_0^1 dy \, X_0(x, y, \Theta) \exp(i \pi x) \sin(\pi y) \in \mathbb{R},
  \label{eq:Phi_zero}
\end{align}
which is unique when a pair of vortices exist in the system, as considered below
(e.g., see Fig.~\ref{fig:4} and Fig.~\ref{fig:5} in Sec.~\ref{subsec:3B}).
When visualizing the temporal phase of the limit-torus in the infinite-dimensional state space,
we project the limit-torus solution $X_0(x - \Phi, y, \Theta)$ onto the $H_{0,2}$-$H_{0,4}$ plane as
\begin{align}
  H_{0,2}(\Theta) &= \int_0^2 dx \int_0^1 dy \, X_0(x - \Phi, y, \Theta) \sin(2 \pi y),
  \\
  H_{0,4}(\Theta) &= \int_0^2 dx \int_0^1 dy \, X_0(x - \Phi, y, \Theta) \sin(4 \pi y),
\end{align}
which are real numbers and depend only on the temporal phase $\Theta$.
When determining the spatial phase of the limit-torus solution,
we introduce the following complex order parameter:
\begin{align}
  A(\Phi, \Theta) \equiv H_{-1,1}(\Theta) \exp(i \pi \Phi)
  = \int_0^2 dx \int_0^1 dy \, X_0(x - \Phi, y, \Theta) \exp(i \pi x) \sin(\pi y),
\end{align}
which corresponds to Eq.~(\ref{eq:Hjk}) with $j = -1$ and $k = 1$.
From Eq.~(\ref{eq:Phi_zero}), the spatial phase $\Phi$ is determined by
\begin{align}
  \Phi = \frac{\arg A(\Phi, \Theta)}{\pi}.
\end{align}

As in $X(x, y, t)$,
considering the boundary conditions
for $Z_{\rm s}(x - \Phi, y, \Theta)$ and $Z_{\rm t}(x - \Phi, y, \Theta)$,
i.e., the periodic boundary condition on $x$ and the Dirichlet zero boundary condition on $y$
as described in Eqs.~(\ref{eq:bcuastx})(\ref{eq:bcuasty}),
we also introduce the following spectral decompositions:
\begin{align}
  Z_{\rm s}(x - \Phi, y, \Theta)
  &= \sum_{j=-\infty}^{\infty} \sum_{k=1}^{\infty} Z_{jk}^{\rm (s)}(\Theta) \exp(-i \pi j \Phi) \exp(i \pi j x) \sin(\pi k y),
  \\
  Z_{\rm t}(x - \Phi, y, \Theta)
  &= \sum_{j=-\infty}^{\infty} \sum_{k=1}^{\infty} Z_{jk}^{\rm (s)}(\Theta) \exp(-i \pi j \Phi) \exp(i \pi j x) \sin(\pi k y),
\end{align}
where the spectral transformations are given by
\begin{align}
  Z_{jk}^{\rm (s)}(\Theta) \exp(-i \pi j \Phi)
  &= \int_0^2 dx \int_0^1 dy \, Z_{\rm s}(x - \Phi, y, \Theta) \exp(-i \pi j x) \sin(\pi k y),
  \\
  Z_{jk}^{\rm (t)}(\Theta) \exp(-i \pi j \Phi)
  &= \int_0^2 dx \int_0^1 dy \, Z_{\rm t}(x - \Phi, y, \Theta) \exp(-i \pi j x) \sin(\pi k y).
\end{align}
The spatial power spectra of the phase sensitivity functions averaged over the temporal phase are defined as
\begin{align}
  P_{\rm s}(j, k)
  &= \frac{1}{2\pi} \int_0^{2\pi} d\Theta \, \left | Z_{jk}^{\rm (s)}(\Theta) \right|^2,
  \label{eq:Ps} \\
  P_{\rm t}(j, k)
  &= \frac{1}{2\pi} \int_0^{2\pi} d\Theta \, \left | Z_{jk}^{\rm (t)}(\Theta) \right|^2.
  \label{eq:Pt}
\end{align}

\subsection{Limit-torus solution and phase sensitivity functions} \label{subsec:3B}

We first summarize our numerical simulations in this paper.
As mentioned in the preceding subsection,
we applied the pseudospectral method,
which is composed of
a Fourier expansion with $256$ modes for the periodic boundary condition on $x$
and a sine expansion with $128$ modes for the Dirichlet zero boundary condition on $y$.
The initial values were prepared
such that the system exhibits oscillatory convection with a pair of vortices.
Because the Rayleigh number was fixed at ${\rm Ra} = 400$,
the traveling velocity and oscillation frequency
were $c = 0$ and $\omega \simeq 532$, respectively.
As mentioned below,
this limit-torus solution possesses reflection symmetry
because the traveling velocity is exactly zero, i.e., $c = 0$.
Here, we note that the theory developed in Sec.~\ref{sec:2} is applicable
for the case of non-zero traveling velocity.

\begin{figure*}
  \begin{center}
    \includegraphics[width=0.5\hsize,clip]{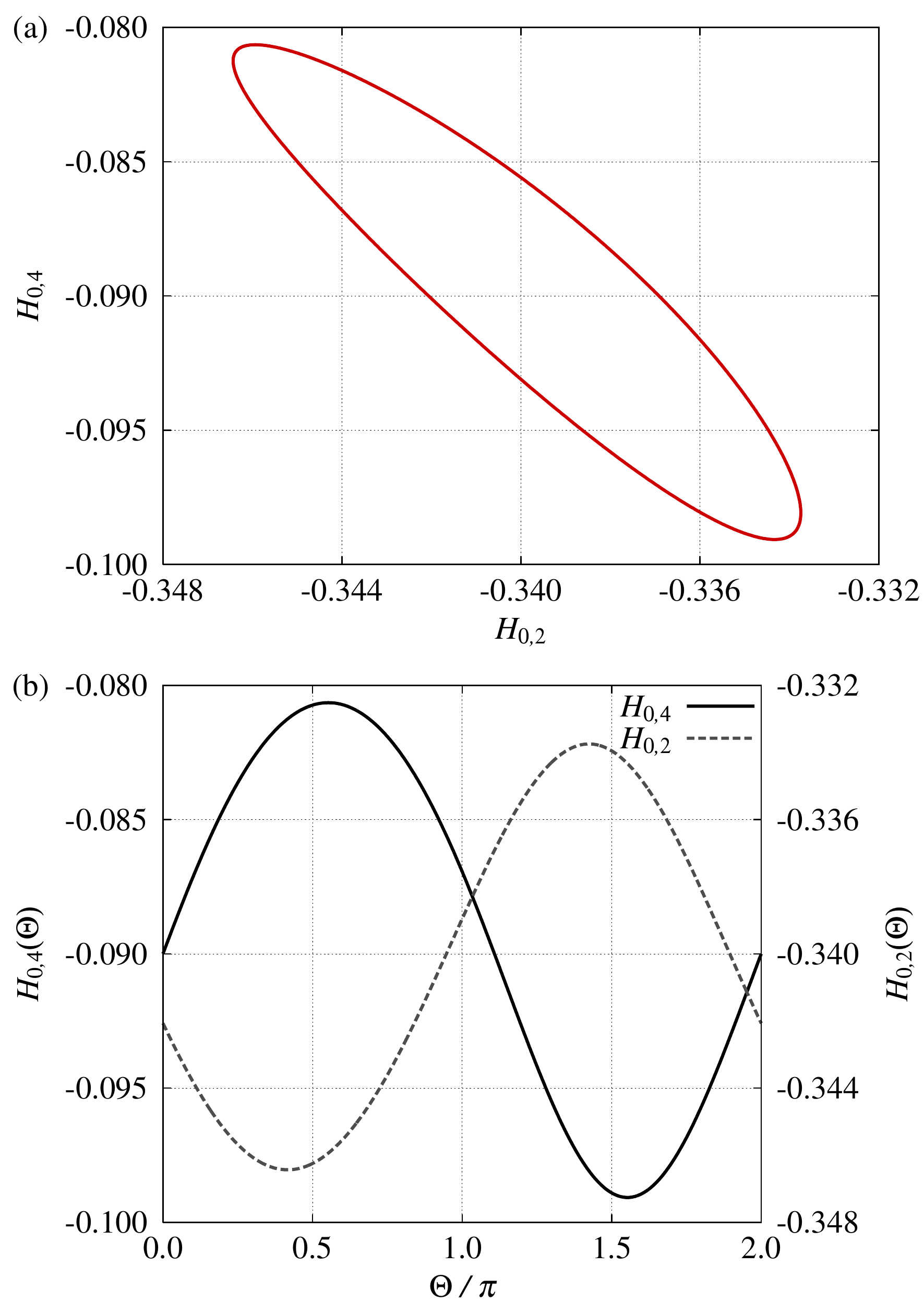}
  \end{center}
  \caption{(Color online)
    (a) Limit-torus orbit projected onto the $H_{0,2}$-$H_{0,4}$ plane.
    (b) Waveforms of $H_{0,2}(\Theta)$ and $H_{0,4}(\Theta)$.
    The Rayleigh number is ${\rm Ra} = 400$; therefore
    the traveling velocity and oscillation frequency
    are $c = 0$ and $\omega \simeq 532$, respectively.
  }
  \label{fig:3}
\end{figure*}

\begin{figure*}
  \begin{center}
    \includegraphics[width=0.5\hsize,clip]{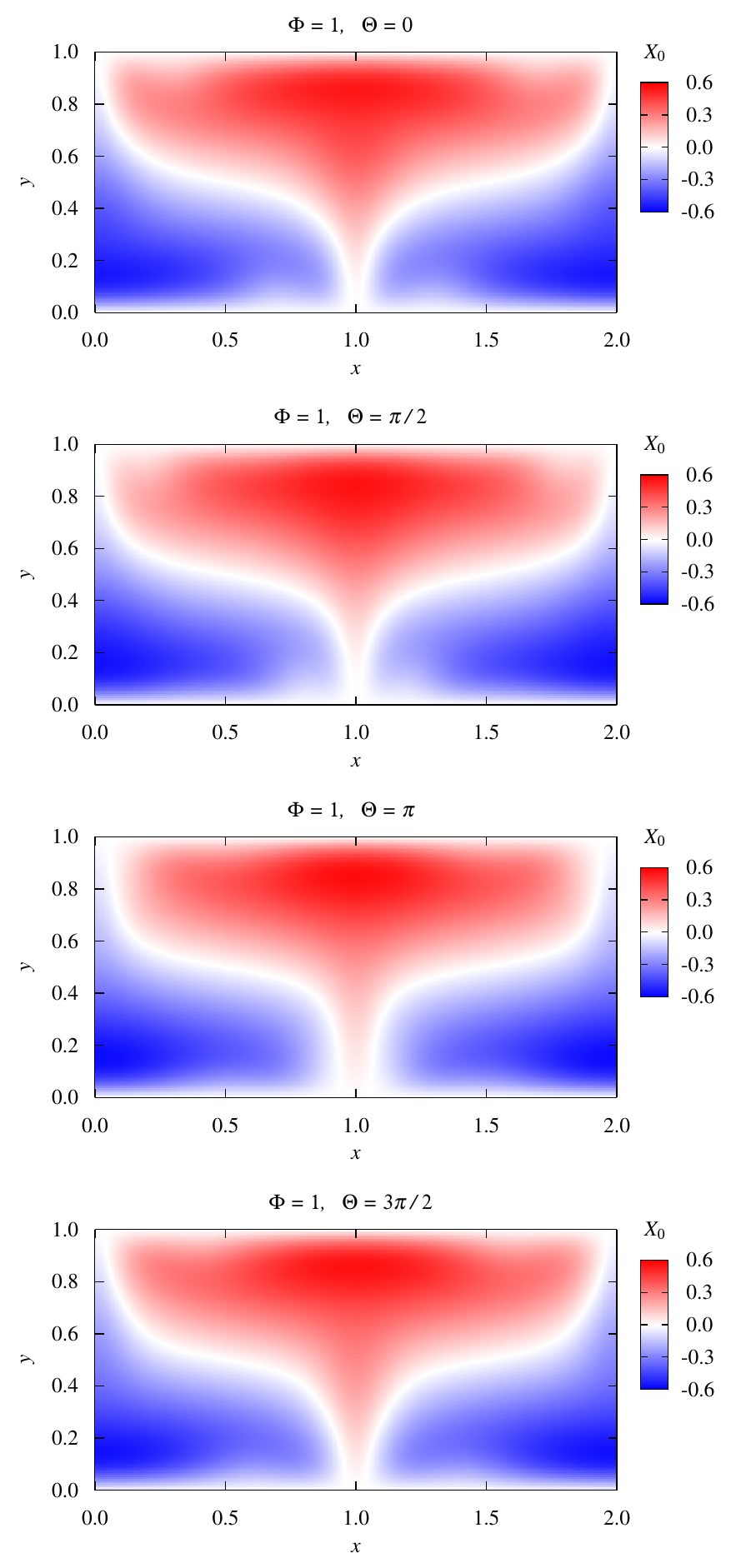}
  \end{center}
  \caption{(Color online)
    Snapshots of $X_0(x - \Phi, y, \Theta)$ with $\Phi = 1$
    for $\Theta = 0$, $\pi/2$, $\pi$, and $3\pi/2$.
  }
  \label{fig:4}
\end{figure*}

\begin{figure*}
  \begin{center}
    \includegraphics[width=1.0\hsize,clip]{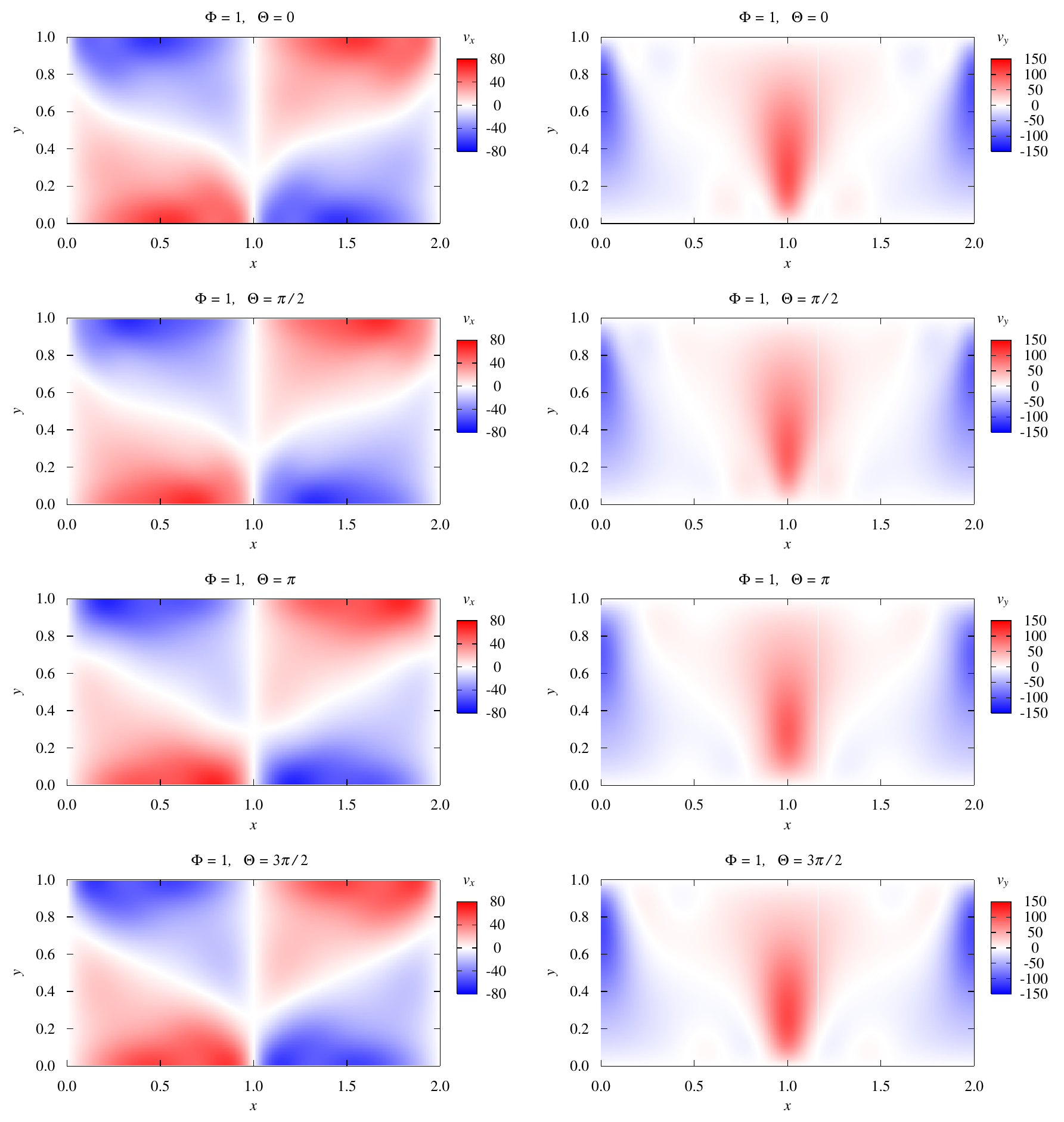}
  \end{center}
  \caption{(Color online)
    Snapshots of both $v_x(x - \Phi, y, \Theta)$
    and $v_y(x - \Phi, y, \Theta)$ with $\Phi = 1$
    for $\Theta = 0$, $\pi/2$, $\pi$, and $3\pi/2$.
  }
  \label{fig:6}
\end{figure*}

\begin{figure*}
  \begin{center}
    \includegraphics[width=1.0\hsize,clip]{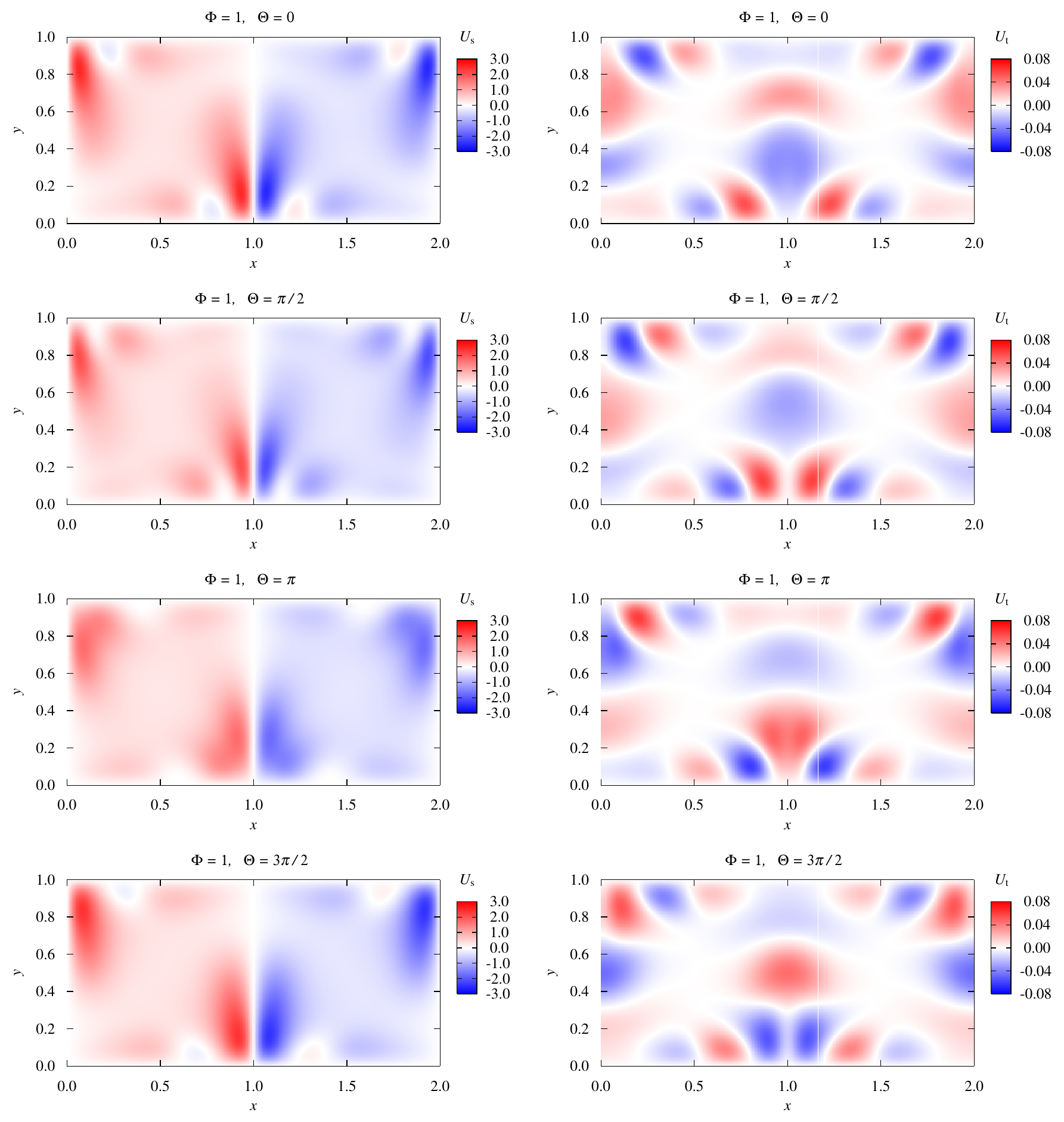}
  \end{center}
  \caption{(Color online)
    Snapshots of both $U_{\rm s}(x - \Phi, y, \Theta)$
    and $U_{\rm t}(x - \Phi, y, \Theta)$ with $\Phi = 1$
    for $\Theta = 0$, $\pi/2$, $\pi$, and $3\pi/2$.
  }
  \label{fig:7}
\end{figure*}

\begin{figure*}
  \begin{center}
    \includegraphics[width=1.0\hsize,clip]{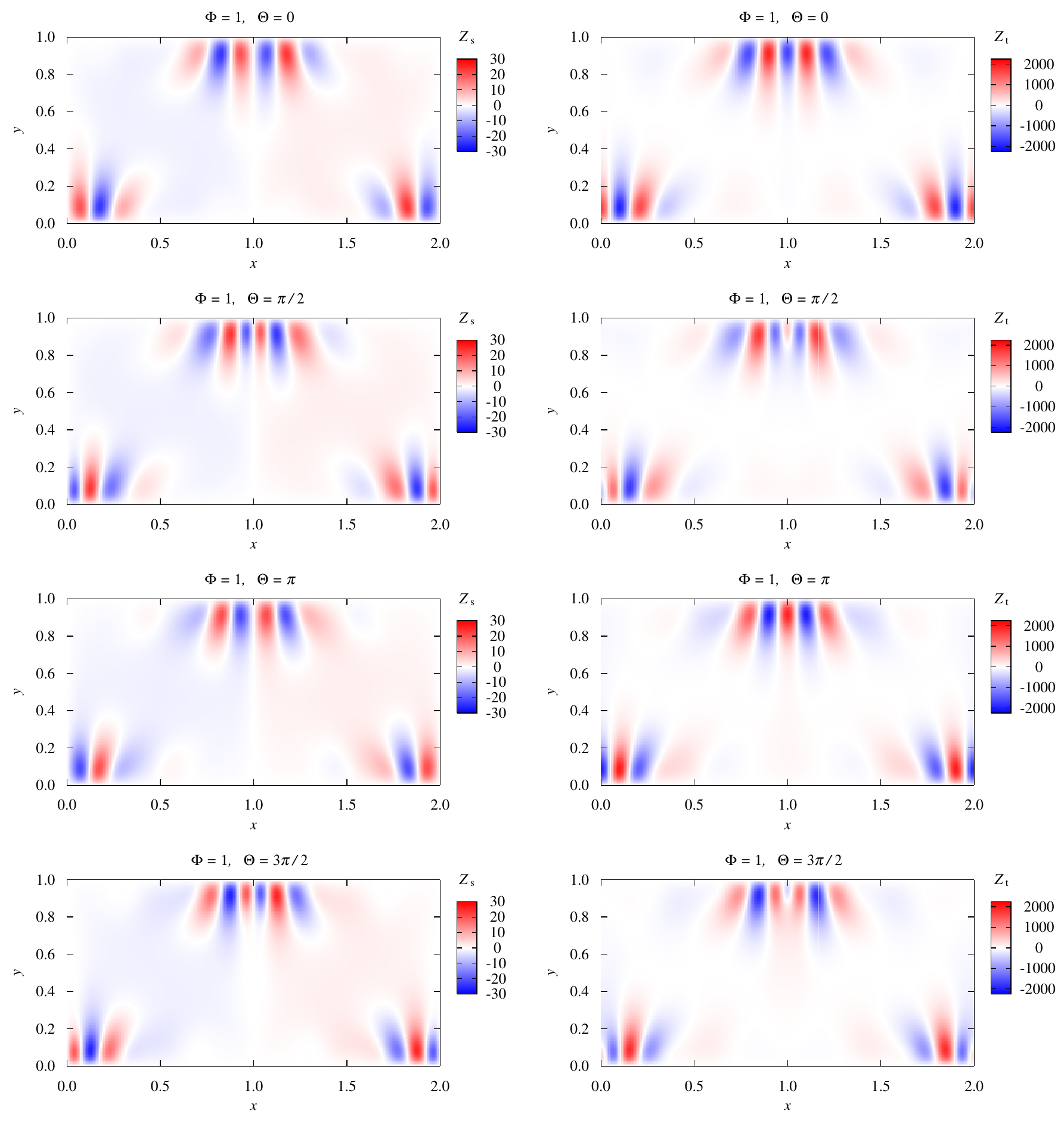}
  \end{center}
  \caption{(Color online)
    Snapshots of both $Z_{\rm s}(x - \Phi, y, \Theta)$
    and $Z_{\rm t}(x - \Phi, y, \Theta)$ with $\Phi = 1$
    for $\Theta = 0$, $\pi/2$, $\pi$, and $3\pi/2$.
  }
  \label{fig:8}
\end{figure*}

Figure~\ref{fig:3} shows the limit-torus orbit projected onto the $H_{0,2}$-$H_{0,4}$ plane,
which was obtained from our numerical simulations of the dynamical equation~(\ref{eq:X}).
Snapshots of the limit-torus solution $X_0(x - \Phi, y, \Theta)$ are shown in Fig.~\ref{fig:4}.
In addition, several quantities associated with the limit-torus solution are shown as follows:
snapshots of the temperature field $T_0(x - \Phi, y, \Theta)$
and the stream function $\psi_0(x - \Phi, y, \Theta)$ are shown in Fig.~\ref{fig:5};
snapshots of the fluid velocity
$\bd{v}_0(x - \Phi, y, \Theta) = ( \partial_y \psi_0(x - \Phi, y, \Theta), \, -\partial_x \psi_0(x - \Phi, y, \Theta) )$
are shown in Fig.~\ref{fig:6};
snapshots of the right zero eigenfunctions,
$U_{\rm s}(x - \Phi, y, \Theta)$ and $U_{\rm t}(x - \Phi, y, \Theta)$,
are shown in Fig.~\ref{fig:7};
snapshots of the phase sensitivity functions,
$Z_{\rm s}(x - \Phi, y, \Theta)$ and $Z_{\rm t}(x - \Phi, y, \Theta)$,
are shown in Fig.~\ref{fig:8}.
We note that the phase sensitivity functions,
$Z_{\rm s}(x - \Phi, y, \Theta)$ and $Z_{\rm t}(x - \Phi, y, \Theta)$,
were obtained using the adjoint method, i.e.,
the relaxation method for Eq.~(\ref{eq:adjoint})
with the orthonormalization condition given by Eq.~(\ref{eq:orthonormal}).

As seen in Fig.~\ref{fig:5},
the spatial phase $\Phi$ can be considered as the ``position'' of the hot plume in the convection,
whereas the temporal phase $\Theta$ represents the ``oscillation'' of the convection.
We also note that the phase sensitivity functions,
$Z_{\rm s}(x - \Phi, y, \Theta)$ and $Z_{\rm t}(x - \Phi, y, \Theta)$,
are spatially localized as seen in Fig.~\ref{fig:8}.
Namely, when the spatial phase is $\Phi = 1$,
the amplitudes of $Z_{\rm s}(x - \Phi, y, \Theta)$ and $Z_{\rm t}(x - \Phi, y, \Theta)$
with respect to the temporal phase $\Theta$
in the bottom-left, bottom-right, and top-center regions of the system
are much larger than those in the other regions;
this fact reflects the dynamics of the spatial pattern of the convective component $X_0(x - \Phi, y, \Theta)$
shown in Fig.~\ref{fig:4}.

Furthermore, in this case,
the limit-torus solution $X_0(x - \Phi, y, \Theta)$
and the phase sensitivity functions, $Z_{\rm s}(x - \Phi, y, \Theta)$ and $Z_{\rm t}(x - \Phi, y, \Theta)$,
possess line symmetry.
As mentioned above, in this case, the traveling velocity is zero, i.e.,
\begin{align}
  c = 0.
  \label{eq:c_zero}
\end{align}
The limit-torus solution representing the oscillatory convection with Eq.~(\ref{eq:c_zero})
possesses the following reflection symmetry (see Fig.~\ref{fig:4}):
\begin{align}
  X_0\bigl( -(x - \Phi), y, \Theta \bigr)
  &= X_0\bigl( x - \Phi, y, \Theta \bigr).
  \label{eq:X0_rs}
\end{align}
From Eq.~(\ref{eq:Us}) and Eq.~(\ref{eq:Ut}),
the right zero eigenfunctions, $U_{\rm s}(x - \Phi, y, \Theta)$ and $U_{\rm t}(x - \Phi, y, \Theta)$,
respectively possess the following reflection anti-symmetry and reflection symmetry (see Fig.~\ref{fig:7}):
\begin{align}
  U_{\rm s}\bigl( -(x - \Phi), y, \Theta \bigr)
  &= -U_{\rm s}\bigl( x - \Phi, y, \Theta \bigr),
  \label{eq:Us_rs} \\
  U_{\rm t}\bigl( -(x - \Phi), y, \Theta \bigr)
  &=  U_{\rm t}\bigl( x - \Phi, y, \Theta \bigr).
  \label{eq:Ut_rs}
\end{align}
Therefore, the phase sensitivity functions, or the left zero eigenfunctions,
also possess the following properties (see Fig.~\ref{fig:8}):
\begin{align}
  Z_{\rm s}\bigl( -(x - \Phi), y, \Theta \bigr)
  &= -Z_{\rm s}\bigl( x - \Phi, y, \Theta \bigr),
  \label{eq:Zs_rs} \\
  Z_{\rm t}\bigl( -(x - \Phi), y, \Theta \bigr)
  &=  Z_{\rm t}\bigl( x - \Phi, y, \Theta \bigr).
  \label{eq:Zt_rs}
\end{align}

Moreover, in this case,
the limit-torus solution $X_0(x - \Phi, y, \Theta)$
and the phase sensitivity functions, $Z_{\rm s}(x - \Phi, y, \Theta)$ and $Z_{\rm t}(x - \Phi, y, \Theta)$,
also possess point symmetry.
For each $\Theta$,
the limit-torus solution and phase sensitivity functions
possess the following properties with respect to a certain point of the system
(see Fig.~\ref{fig:4} and Fig.~\ref{fig:8}):
\begin{align}
  X_0(-x_\delta, -y_\delta, \Theta)
  &= -X_0(x_\delta, y_\delta, \Theta),
  \label{eq:X0_ps}
  \\
  Z_{\rm s}(-x_\delta, -y_\delta, \Theta)
  &=  Z_{\rm s}(x_\delta, y_\delta, \Theta),
  \label{eq:Zs_ps}
  \\
  Z_{\rm t}(-x_\delta, -y_\delta, \Theta)
  &= -Z_{\rm t}(x_\delta, y_\delta, \Theta),
  \label{eq:Zt_ps}
\end{align}
where $x_\delta = x - \Phi/2$ and $y_\delta = y - 1/2$.

\subsection{Effective phase sensitivity functions} \label{subsec:3C}

\begin{figure*}
  \begin{center}
    \includegraphics[width=0.5\hsize,clip]{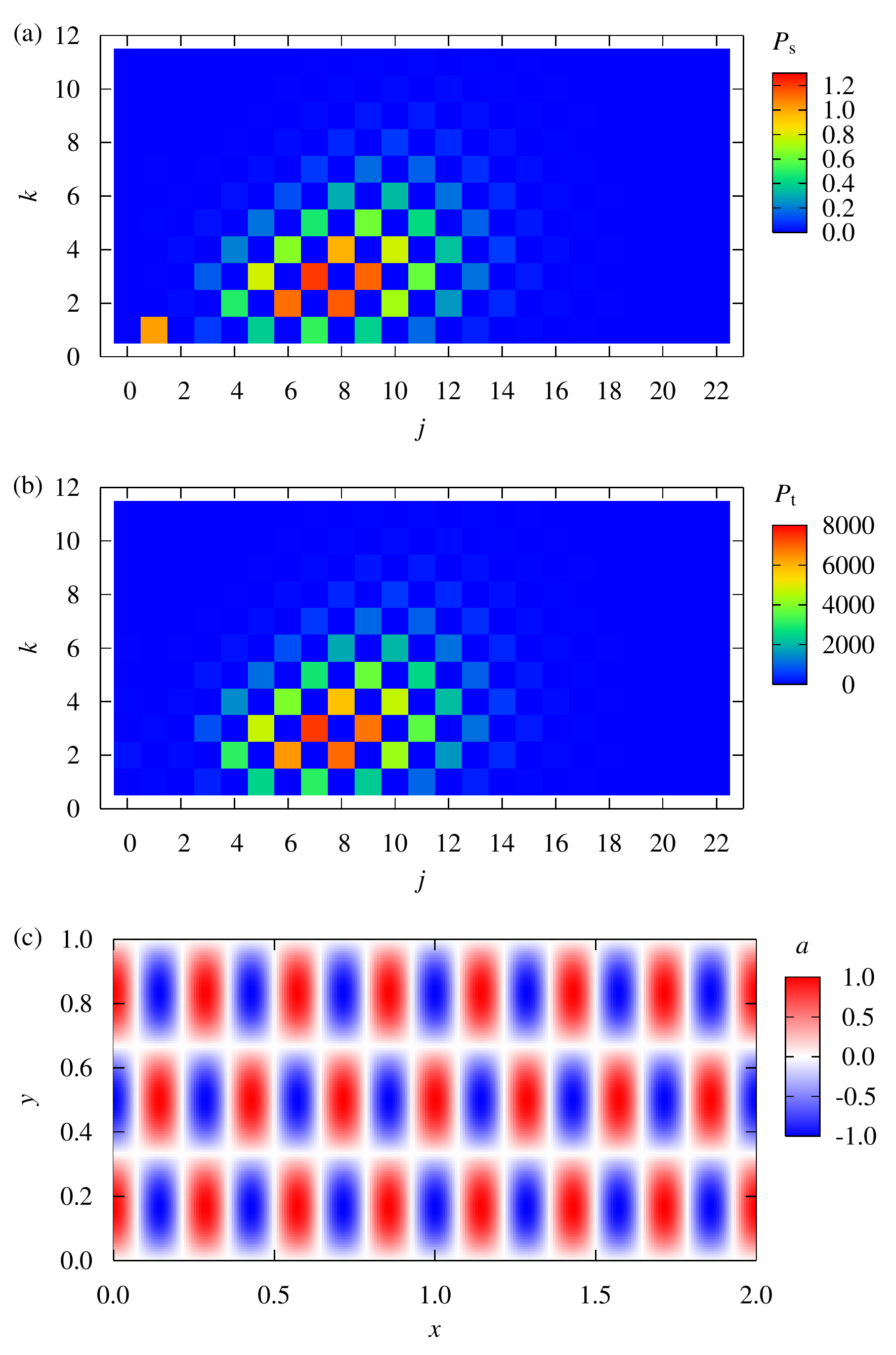}
  \end{center}
  \caption{(Color online)
    (a) Spatial power spectrum of $Z_{\rm s}(x - \Phi, y, \Theta)$ averaged over $\Theta$,
    i.e., $P_{\rm s}(j, k)$.
    (b) Spatial power spectrum of $Z_{\rm t}(x - \Phi, y, \Theta)$ averaged over $\Theta$,
    i.e., $P_{\rm t}(j, k)$.
    (c) Spatial pattern $a(x, y) = \cos(\pi j x) \sin(\pi k y)$ with $j = 7$ and $k = 3$.
  }
  \label{fig:9}
\end{figure*}

\begin{figure*}
  \begin{center}
    \includegraphics[width=0.5\hsize,clip]{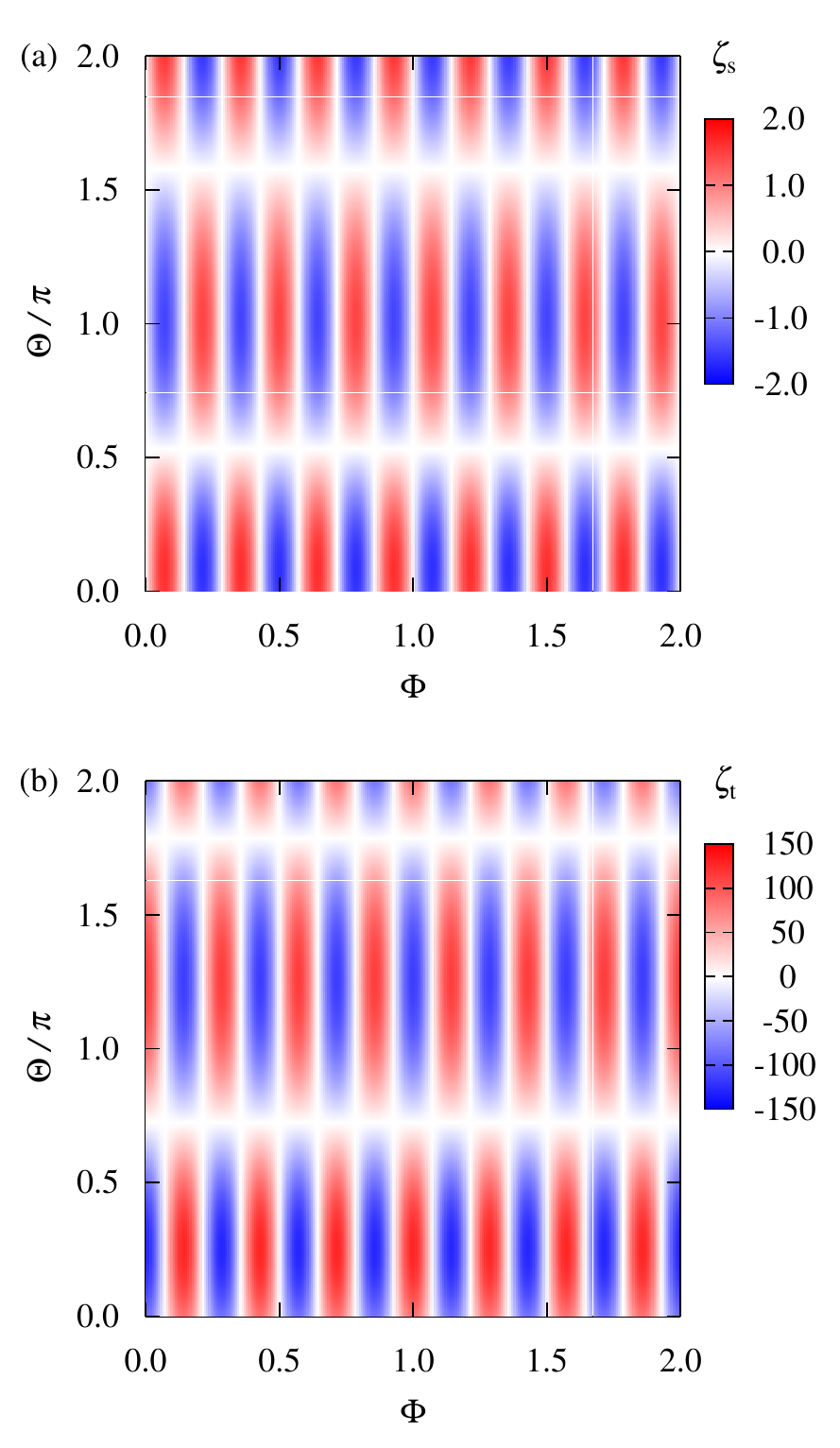}
  \end{center}
  \caption{(Color online)
    (a) Effective phase sensitivity function for the spatial phase,
    $\zeta_{\rm s}(\Phi, \Theta)$.
    (b) Effective phase sensitivity function for the temporal phase,
    $\zeta_{\rm t}(\Phi, \Theta)$.
    The spatial pattern $a(x, y)$ of the perturbation
    is given by Eq.~(\ref{eq:a_cs})
    and
    is shown in Fig.~\ref{fig:9}(c).
  }
  \label{fig:10}
\end{figure*}

\begin{figure*}
  \begin{center}
    \includegraphics[width=0.5\hsize,clip]{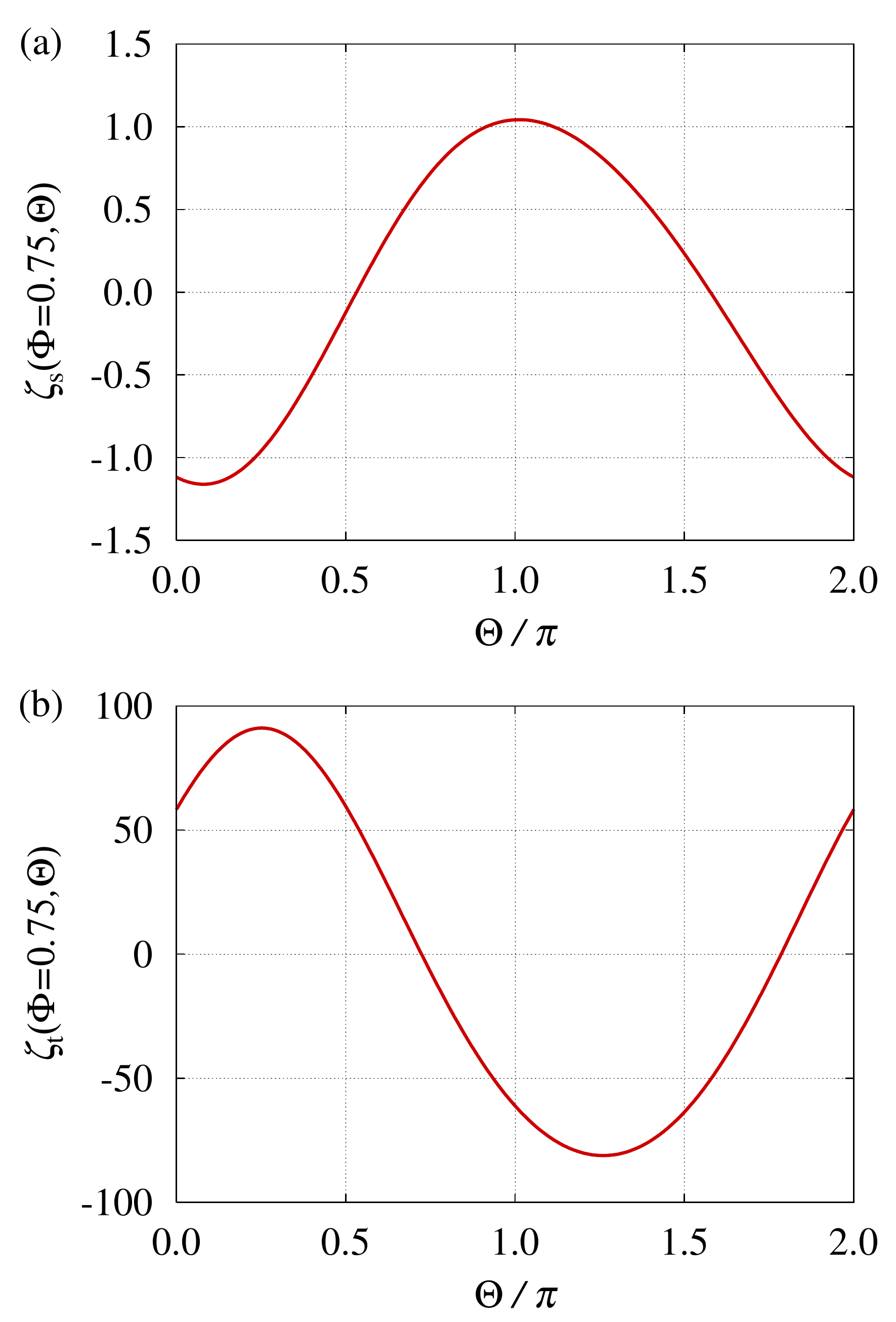}
  \end{center}
  \caption{(Color online)
    (a) Effective phase sensitivity function for the spatial phase, $\zeta_{\rm s}(\Phi, \Theta)$,
    plotted as a function of $\Theta$ with $\Phi = 0.75$.
    (b) Effective phase sensitivity function for the temporal phase, $\zeta_{\rm t}(\Phi, \Theta)$,
    plotted as a function of $\Theta$ with $\Phi = 0.75$.
  }
  \label{fig:11}
\end{figure*}

In this subsection,
we calculate the effective phase sensitivity functions obtained in Sec.~\ref{subsec:2E}.
Before illustrating the effective phase sensitivity functions,
the spatial power spectra of the phase sensitivity functions averaged over the temporal phase
as defined in Eq.~(\ref{eq:Ps}) and Eq.~(\ref{eq:Pt}),
i.e., $P_{\rm s}(j, k)$ and $P_{\rm t}(j, k)$, 
are shown in Fig.~\ref{fig:9}(a) and Fig.~\ref{fig:9}(b), respectively.
Owing to the point-symmetry of the phase sensitivity functions,
given in Eqs.~(\ref{eq:Zs_ps})(\ref{eq:Zt_ps}),
both $P_{\rm s}(j, k)$ and $P_{\rm t}(j, k)$ exhibit checkerboard patterns.
The power of the mode $(j, k) = (7, 3)$ is the largest for $P_{\rm s}(j, k)$ and $P_{\rm t}(j, k)$.
To illustrate the effective phase sensitivity functions,
we consider a corresponding spatial pattern
\begin{align}
  a(x, y) = \cos(\pi j x) \sin(\pi k y),
  \label{eq:a_cs}
\end{align}
where $j = 7$ and $k = 3$.
Figure~\ref{fig:9}(c) shows the spatial pattern $a(x, y)$,
for which the effective phase sensitivity functions
are shown in Fig.~\ref{fig:10} with respect to $\Phi$ and $\Theta$.
In addition, the effective phase sensitivity functions with $\Phi = 0.75$
are shown in Fig.~\ref{fig:11} as a function of $\Theta$.

As seen in Fig.~\ref{fig:10} and Fig.~\ref{fig:11},
the effective phase sensitivity functions,
$\zeta_{\rm s}(\Phi, \Theta)$ and $\zeta_{\rm t}(\Phi, \Theta)$,
exhibit both positive and negative values.
Namely, when the effective phase sensitivity function for the spatial phase,
$\zeta_{\rm s}(\Phi, \Theta)$, is positive (negative),
the spatial phase $\Phi$ is advanced (delayed) by applying a positive perturbation.
Similarly, when the effective phase sensitivity function for the temporal phase,
$\zeta_{\rm t}(\Phi, \Theta)$, is positive (negative),
the temporal phase $\Theta$ is advanced (delayed) by applying a positive perturbation.

\subsection{Phase coupling functions} \label{subsec:3D}

In this subsection,
we calculate the phase coupling functions obtained in Sec.~\ref{subsec:2F}.
The anti-symmetric components of the phase coupling functions,
$\Gamma_{\rm s}^{\rm (a)}(\varDelta\Phi, \varDelta\Theta)$
and
$\Gamma_{\rm t}^{\rm (a)}(\varDelta\Phi, \varDelta\Theta)$,
are shown in Fig.~\ref{fig:12}(a) and Fig.~\ref{fig:12}(b), respectively.
As shown in Eqs.~(\ref{eq:Gsa_as})(\ref{eq:Gta_as}),
both $\Gamma_{\rm s}^{\rm (a)}(\varDelta\Phi, \varDelta\Theta)$
and $\Gamma_{\rm t}^{\rm (a)}(\varDelta\Phi, \varDelta\Theta)$
possess anti-symmetry with respect to the origin, i.e.,
$\varDelta\Phi = \varDelta\Theta = 0$.

Here, we describe another symmetry of
the anti-symmetric components of the phase coupling functions in this reflection symmetric case.
The phase coupling functions given by Eqs.~(\ref{eq:Gamma_s})(\ref{eq:Gamma_t}) are written in the following forms:
\begin{align}
  \Gamma_{\rm s}(\varDelta\Phi, \varDelta\Theta)
  &= \frac{1}{2\pi} \int_0^{2\pi} d\lambda \int_0^2 dx \int_0^1 dy \, Z_{\rm s}(x - \varDelta\Phi, y, \lambda + \varDelta\Theta)
  \Bigl[ X_0(x, y, \lambda) - X_0(x - \varDelta\Phi, y, \lambda + \varDelta\Theta) \Bigr],
  \\
  \Gamma_{\rm t}(\varDelta\Phi, \varDelta\Theta)
  &= \frac{1}{2\pi} \int_0^{2\pi} d\lambda \int_0^2 dx \int_0^1 dy \, Z_{\rm t}(x - \varDelta\Phi, y, \lambda + \varDelta\Theta)
  \Bigl[ X_0(x, y, \lambda) - X_0(x - \varDelta\Phi, y, \lambda + \varDelta\Theta) \Bigr].
\end{align}
From Eqs.~(\ref{eq:X0_rs})(\ref{eq:Zs_rs})(\ref{eq:Zt_rs}),
the phase coupling functions,
$\Gamma_{\rm s}(\varDelta\Phi, \varDelta\Theta)$ and $\Gamma_{\rm t}(\varDelta\Phi, \varDelta\Theta)$,
respectively possess the following reflection anti-symmetry and reflection symmetry:
\begin{align}
  \Gamma_{\rm s}(-\varDelta\Phi, \varDelta\Theta)
  &= -\Gamma_{\rm s}(\varDelta\Phi, \varDelta\Theta),
  \label{eq:Gs_rs} \\
  \Gamma_{\rm t}(-\varDelta\Phi, \varDelta\Theta)
  &=  \Gamma_{\rm t}(\varDelta\Phi, \varDelta\Theta).
  \label{eq:Gt_rs}
\end{align}
Therefore, the anti-symmetric components of the phase coupling functions,
$\Gamma_{\rm s}^{\rm (a)}(\varDelta\Phi, \varDelta\Theta)$ and $\Gamma_{\rm t}^{\rm (a)}(\varDelta\Phi, \varDelta\Theta)$,
also possess the following properties (see Fig.~\ref{fig:12}):
\begin{align}
  \Gamma_{\rm s}^{\rm (a)}(-\varDelta\Phi, \varDelta\Theta)
  &= -\Gamma_{\rm s}^{\rm (a)}(\varDelta\Phi, \varDelta\Theta),
  \label{eq:Gsa_rs} \\
  \Gamma_{\rm t}^{\rm (a)}(-\varDelta\Phi, \varDelta\Theta)
  &=  \Gamma_{\rm t}^{\rm (a)}(\varDelta\Phi, \varDelta\Theta).
  \label{eq:Gta_rs}
\end{align}
In addition,
considering the symmetries given by Eqs.~(\ref{eq:Gsa_as})(\ref{eq:Gta_as}) and Eqs.~(\ref{eq:Gsa_rs})(\ref{eq:Gta_rs}),
we also obtain the following properties (see Fig.~\ref{fig:12}):
\begin{align}
  &  \Gamma_{\rm s}^{\rm (a)}( \varDelta\Phi, -\varDelta\Theta)
  =  \Gamma_{\rm s}^{\rm (a)}( \varDelta\Phi,  \varDelta\Theta),
  \\
  &  \Gamma_{\rm t}^{\rm (a)}( \varDelta\Phi, -\varDelta\Theta)
  = -\Gamma_{\rm t}^{\rm (a)}( \varDelta\Phi,  \varDelta\Theta).
\end{align}

From the above symmetries, in this case,
it is sufficient to investigate only the following region: $(\varDelta \Phi, \varDelta \Theta / \pi) \in [0, 1] \times [0, 1]$.
Figure~\ref{fig:12}(c) shows the nullclines, fixed points, and typical orbits of the phase differences in this region.
The typical orbits were obtained from the numerical simulations of Eqs.~(\ref{eq:DPhi})(\ref{eq:DTheta}).
As seen in Fig.~\ref{fig:12}(c),
the spatial and temporal in-phase state, i.e.,
$(\varDelta \Phi, \varDelta \Theta / \pi) = (0, 0)$,
is globally stable under the phase reduction approximation.
The spatial phase difference $\varDelta \Phi$ monotonously decreases to zero,
whereas the temporal phase difference $\varDelta \Theta$ first oscillates and then becomes zero.

\section{Comparisons with direct numerical simulations} \label{sec:4}

In this section,
we compare the theoretical values obtained in Sec.~\ref{sec:3}
with direct numerical simulations of oscillatory convection.

\subsection{Spatiotemporal phase responses of oscillatory convection to weak impulses}

\begin{figure*}
  \begin{center}
    \includegraphics[width=0.5\hsize,clip]{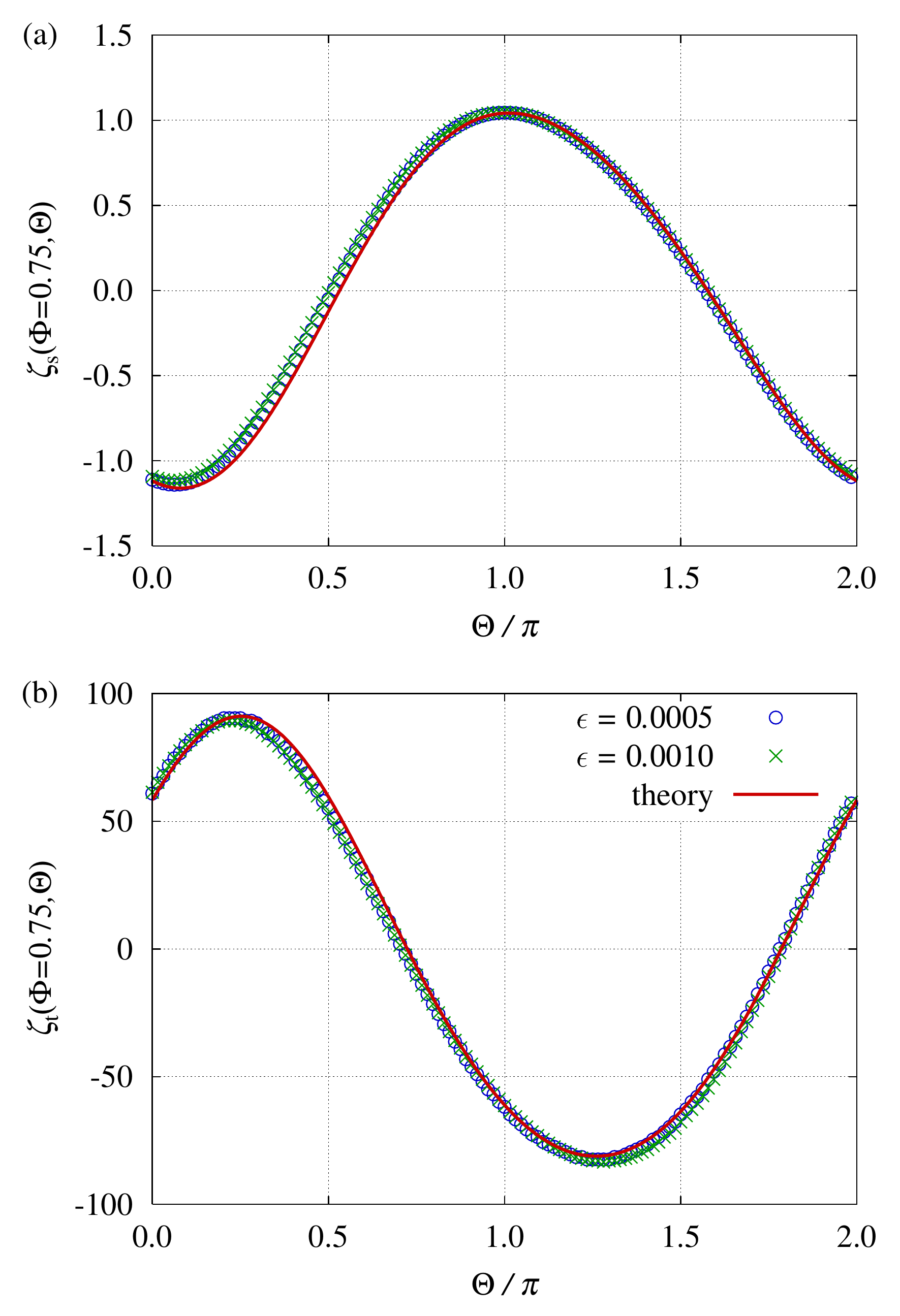}
  \end{center}
  \caption{(Color online)
    Comparisons of the effective phase sensitivity functions, i.e.,
    (a) $\zeta_{\rm s}(\Phi = 0.75, \Theta)$ and
    (b) $\zeta_{\rm t}(\Phi = 0.75, \Theta)$,
    between direct numerical simulations with impulse intensity $\epsilon$
    and the theoretical curves (theory).
  }
  \label{fig:13}
\end{figure*}

In this subsection,
we make a comparison of the effective phase sensitivity functions
between the theoretical values obtained in Sec.~\ref{subsec:3C}, i.e., Fig.~\ref{fig:11},
and direct numerical simulations of oscillatory convection with weak impulses
described by Eq.~(\ref{eq:Xp}) using Eqs.~(\ref{eq:p_aq})(\ref{eq:a_cs}).
The comparison of the effective phase sensitivity functions,
$\zeta_{\rm s}(\Phi = 0.75, \Theta)$ and $\zeta_{\rm t}(\Phi = 0.75, \Theta)$,
between the direct numerical simulations with impulse intensity $\epsilon$ and the theoretical curves
are shown in Fig.~\ref{fig:13}.
The simulation results agree quantitatively with the theory.

\subsection{Spatiotemporal phase synchronization between weakly coupled systems of oscillatory convection}

\begin{figure*}
  \begin{center}
    \includegraphics[width=0.5\hsize,clip]{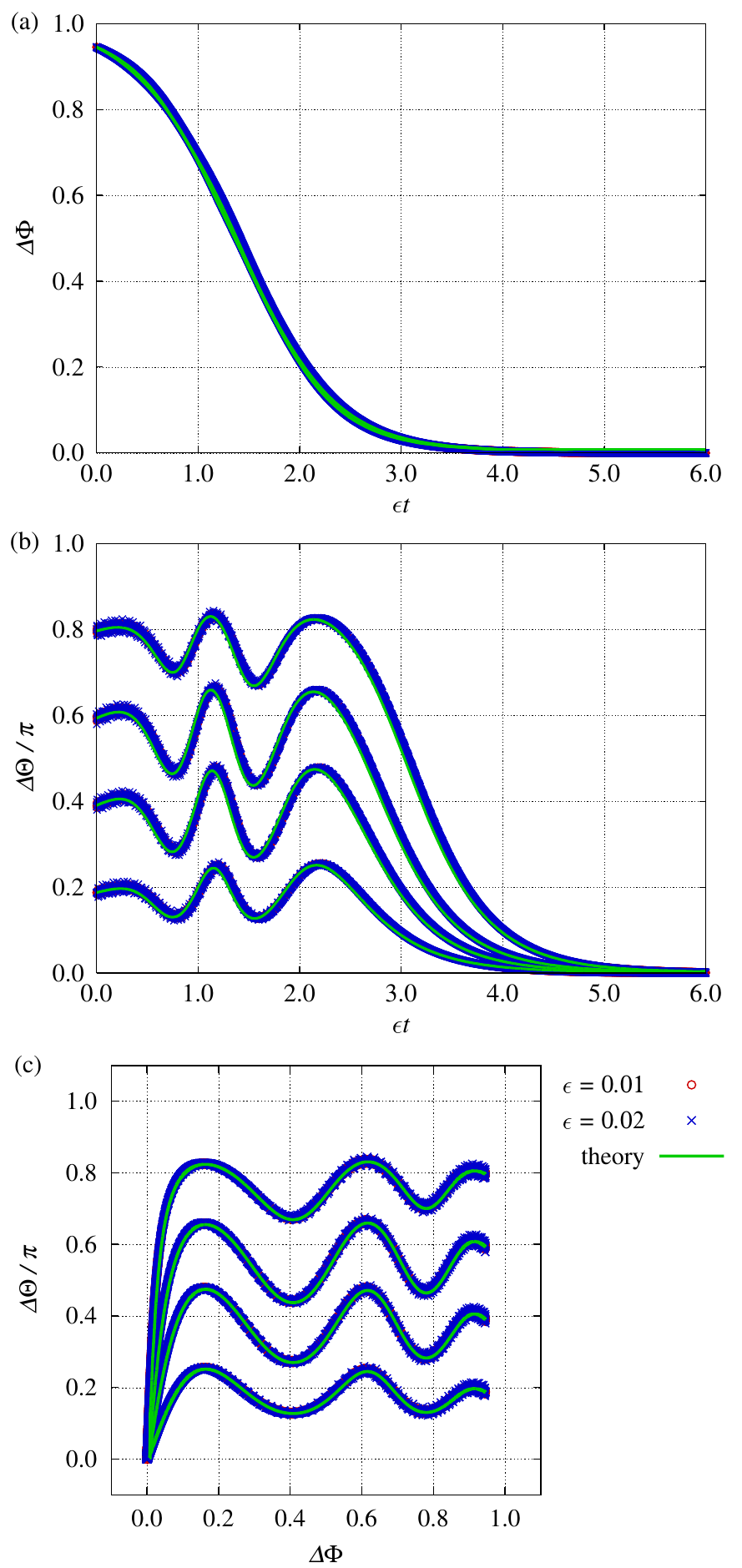}
  \end{center}
  \caption{(Color online)
    Comparisons of the time evolution of the phase differences
    between direct numerical simulations with coupling intensity $\epsilon$
    and the theoretical curves (theory).
    (a) $\varDelta \Phi$ vs. $\epsilon t$.
    (b) $\varDelta \Theta / \pi$ vs. $\epsilon t$.
    (c) $\varDelta \Theta / \pi$ vs. $\varDelta \Phi$.
    The initial values are
    $(\varDelta \Phi, \varDelta \Theta / \pi) = (0.95, 0.20)$,
    $(0.95, 0.40)$, $(0.95, 0.60)$, and $(0.95, 0.80)$.
  }
  \label{fig:14}
\end{figure*}

In this subsection,
we make a comparison on time evolution of phase differences
between the theoretical values obtained in Sec.~\ref{subsec:3D}, i.e., Fig.~\ref{fig:12}(c),
and direct numerical simulations of two weakly coupled systems of oscillatory convection
described by Eq.~(\ref{eq:Xsigma}).
The comparison of the time evolution of the phase differences
between the direct numerical simulations with coupling intensity $\epsilon$ and the theoretical curves
are shown in Fig.~\ref{fig:14}.
The simulation results agree quantitatively with the theory.

\section{Concluding remarks} \label{sec:5}

Our investigations in this paper are summarized as follows.
In Sec.~\ref{sec:2},
we formulated the theory for the phase description of oscillatory convection with a spatially translational mode.
In particular, we derived the phase sensitivity functions for the spatial and temporal phases.
Details of the derivation of the adjoint operator, which provides the phase sensitivity functions,
are given in App.~\ref{sec:A}.
Treatments of the boundary forcing are described in App.~\ref{sec:B}.
We also derived the phase coupling functions for the spatial and temporal phases.
In Sec.~\ref{sec:3},
we illustrated the theory using the numerical analysis of the oscillatory convection.
In particular, we obtained the phase sensitivity functions and phase coupling functions.
In Sec.~\ref{sec:4},
we made comparisons between the theory and direct numerical simulations:
the spatiotemporal phase responses of oscillatory convection to weak impulses
and
the spatiotemporal phase synchronization between two weakly coupled systems of oscillatory convection.
The theoretical predictions were successfully confirmed by the direct numerical simulations.

Here, we give some concluding remarks.
First, we summarize three types of solutions
to partial differential equations for the field $X(x, t)$
and the corresponding phase description methods considered so far~\footnote{
  We do not consider the neighborhood of the drift bifurcation point,
  in which the corresponding critical mode should be further taken into account
  (see Refs.~\cite{ref:ohta01,ref:ei02} for this point
  in the case of traveling pulse solutions to reaction-diffusion equations).
}.
(A) a traveling solution:
$X(x, t) = X_0(x - \Phi(t))$ with $\dot{\Phi}(t) = c$.
(B) an oscillating solution:
$X(x, t) = X_0(x, \Theta(t))$ with $\dot{\Theta}(t) = \omega$.
(C) a traveling and oscillating solution:
$X(x, t) = X_0(x - \Phi(t), \Theta(t))$ with $\dot{\Phi}(t) = c$ and $\dot{\Theta}(t) = \omega$.
The phase description method for Type~(A) has been developed
in Refs.~\cite{ref:kawamura07,ref:kawamura08,ref:kawamura10}
(see also Refs.~\cite{ref:jalics10,ref:kilpatrick12,ref:lober12});
this method is closely related to the phase reduction approach to spatially periodic patterns
in that the phase is associated with spatial translational symmetry breaking
\cite{ref:pomeau79,ref:cross83,ref:cross84,ref:brand83,ref:brand84,ref:fauve87}
(see also Refs.~\cite{ref:kuramoto84,ref:manneville90,ref:mori97,ref:cross93,ref:cross09,ref:kuramoto84a,ref:kuramoto89,ref:ohta87}).
As mentioned in Refs.~\cite{ref:kawamura11,ref:kawamura13,ref:nakao12},
Type~(A) can be considered as a special case of Type~(B).
Type~(B) can be considered as a limit-cycle solution and possesses only one phase variable,
whereas
Type~(C) can be considered as a limit-torus solution and possesses two phase variables.
The phase description method developed in this paper belongs to Type~(C),
and it can be considered as a generalization of our phase description method for Type~(B)
developed in Refs.~\cite{ref:kawamura11,ref:kawamura13,ref:nakao12}.

Second, we note the spatial reflection symmetry of spatial patterns.
Consider the pattern formation in a system of spatial reflection symmetry:
when a spatial pattern does not break the reflection symmetry,
the traveling velocity is zero, $c = 0$;
meanwhile,
when a spatial pattern does break the reflection symmetry,
the traveling velocity becomes non-zero, $c \ne 0$.
The traveling velocity of the oscillatory cylindrical-Hele-Shaw convection is zero;
however, the phase description method itself
is applicable for the case of non-zero traveling velocity.
If the reflection symmetry of limit-torus solution, i.e., Eq.~(\ref{eq:X0_rs}), is lost,
then those of the phase sensitivity functions,
i.e., Eqs.~(\ref{eq:Zs_rs})(\ref{eq:Zt_rs}), are lost,
and those of the phase coupling functions,
i.e., Eqs.~(\ref{eq:Gs_rs})(\ref{eq:Gt_rs}) and Eqs.~(\ref{eq:Gsa_rs})(\ref{eq:Gta_rs}), are also lost.

Third, we note ubiquitousness of the limit-torus solutions to partial differential equations.
A limit-torus solution to an ordinary differential equation represents a quasi-periodic oscillator,
and a phase description method for the quasi-periodic oscillator has also been developed
\cite{ref:izhikevich99,ref:demir10}.
However, there have been few studies on the synchronization of quasi-periodic oscillators;
this may be due to the fact that limit-cycle or chaotic oscillations are ubiquitous,
but quasi-periodic oscillations are rather rare in ordinary differential equations.
In contrast, a limit-torus solution is ubiquitous
in a partial differential equation that possesses some spatial translational symmetry.
From this point of view,
a systematic analysis of a set of phase equations, such as
Eqs.~(\ref{eq:Phi_q})(\ref{eq:Theta_q}),
Eqs.~(\ref{eq:Phi_Gamma})(\ref{eq:Theta_Gamma}), and
Eqs.~(\ref{eq:DPhi})(\ref{eq:DTheta}),
is meaningful and important.
It should also be noted that
these phase equations are universal and invariant for limit-torus solutions.

Finally, we remark the broad applicability of our phase description approach,
which is not restricted to oscillatory cylindrical-Hele-Shaw convection.
The phase description method can be generalized to
traveling and oscillating localized convection (i.e., a traveling breather)
in a binary fluid system (see, e.g., Ref.~\cite{ref:watanabe12})
or
traveling and oscillating convection
in a rotating fluid annulus system (see, e.g., Ref.~\cite{ref:read06}).
Similar phase description methods can also be developed for
oscillating spots of zero traveling velocity (see, e.g., Ref.~\cite{ref:hagberg94})
or
traveling breathers of non-zero traveling velocity (see, e.g., Ref.~\cite{ref:yadome11})
in reaction-diffusion systems.

\begin{acknowledgments}
  Y.K. is grateful to members of both
  the Earth Evolution Modeling Research Team and
  the Nonlinear Dynamics and Its Application Research Team
  at IFREE/JAMSTEC for fruitful comments.
  Y.K. is also grateful for financial support by
  JSPS KAKENHI Grant Number 25800222.
  H.N. is grateful for financial support by
  JSPS KAKENHI Grant Numbers 25540108 and 22684020,
  CREST Kokubu project of JST, and
  FIRST Aihara project of JSPS.
\end{acknowledgments}

\appendix

\section{Derivation of the adjoint operator} \label{sec:A}

In this appendix,
we describe the details of the derivation of the adjoint operator ${\cal L}^\ast(x - \Phi, y, \Theta)$
given in Eqs.~(\ref{eq:calLast})(\ref{eq:Last})
(see also, e.g., Refs.~\cite{ref:zwillinger98,ref:keener00} for mathematical terms).
The derivation procedure is similar to that performed in Ref.~\cite{ref:kawamura13}.
From Eqs.~(\ref{eq:calL})(\ref{eq:L}),
the linear operator ${\cal L}(x - \Phi, y, \Theta)$ is given by the following form:
\begin{align}
  {\cal L}(x - \Phi, y, \Theta) u(x - \Phi, y, \Theta)
  = \frac{\partial^2 u}{\partial x^2}
  + \frac{\partial^2 u}{\partial y^2}
  - \frac{\partial \psi_0}{\partial y} \frac{\partial u}{\partial x}
  + \frac{\partial \psi_0}{\partial x} \frac{\partial u}{\partial y}
  + \frac{\partial \psi_u}{\partial x} \left( \frac{\partial X_0}{\partial y} - 1 \right)
  - \frac{\partial \psi_u}{\partial y} \frac{\partial X_0}{\partial x}
  + c \frac{\partial u}{\partial x}
  - \omega \frac{\partial u}{\partial \Theta}.
\end{align}
By partial integration,
each term of the inner product
$\pd{u^\ast(x - \Phi, y, \Theta), {\cal L}(x - \Phi, y, \Theta) u(x - \Phi, y, \Theta)}$
is transformed into
\begin{align}
  \PD{ u^\ast, \frac{\partial^2 u}{\partial x^2} }
  &= \frac{1}{2\pi} \int_0^{2\pi} d\Theta \int_0^1 dy \,
  \left\{ \left[ u^\ast \, \frac{\partial u}{\partial x} \right]_{x=0}^{x=2}
  - \left[ \frac{\partial u^\ast}{\partial x} \, u \right]_{x=0}^{x=2} \right\}
  + \PD{ \frac{\partial^2 u^\ast}{\partial x^2}, u },
  \\
  \PD{ u^\ast, \frac{\partial^2 u}{\partial y^2} }
  &= \frac{1}{2\pi} \int_0^{2\pi} d\Theta \int_0^2 dx \,
  \left\{ \left[ u^\ast \, \frac{\partial u}{\partial y} \right]_{y=0}^{y=1}
  - \left[ \frac{\partial u^\ast}{\partial y} \, u \right]_{y=0}^{y=1} \right\}
  + \PD{ \frac{\partial^2 u^\ast}{\partial y^2}, u },
  \\
  \PD{ u^\ast, - \frac{\partial \psi_0}{\partial y} \frac{\partial u}{\partial x} }
  &= -\frac{1}{2\pi} \int_0^{2\pi} d\Theta \int_0^1 dy \,
  \left[ u^\ast \, \frac{\partial \psi_0}{\partial y} \, u \right]_{x=0}^{x=2}
  + \PD{ \frac{\partial}{\partial x} \left[ u^\ast \frac{\partial \psi_0}{\partial y} \right] , u },
  \\
  \PD{ u^\ast, \frac{\partial \psi_0}{\partial x} \frac{\partial u}{\partial y} }
  &= \frac{1}{2\pi} \int_0^{2\pi} d\Theta \int_0^2 dx \,
  \left[ u^\ast \, \frac{\partial \psi_0}{\partial x} \, u \right]_{y=0}^{y=1}
  + \PD{ -\frac{\partial}{\partial y} \left[ u^\ast \frac{\partial \psi_0}{\partial y} \right] , u },
  \\
  \PD{ u^\ast, \frac{\partial \psi_u}{\partial x} \left( \frac{\partial X_0}{\partial y} - 1 \right) }
  &= \frac{1}{2\pi} \int_0^{2\pi} d\Theta \int_0^1 dy \,
  \left[ u^\ast \, \left( \frac{\partial X_0}{\partial y} - 1 \right) \, \psi_u \right]_{x=0}^{x=2}
  + \PD{ -\frac{\partial}{\partial x} \left[ u^\ast \left( \frac{\partial X_0}{\partial y} - 1 \right) \right] , \psi_u },
  \label{eq:PD_P_u_1} \\
  \PD{ u^\ast, -\frac{\partial \psi_u}{\partial y} \frac{\partial X_0}{\partial x} }
  &= -\frac{1}{2\pi} \int_0^{2\pi} d\Theta \int_0^2 dx \,
  \left[ u^\ast \, \frac{\partial X_0}{\partial x} \, \psi_u \right]_{y=0}^{y=1}
  + \PD{ \frac{\partial}{\partial y} \left[ u^\ast \frac{\partial X_0}{\partial x} \right] , \psi_u },
  \label{eq:PD_P_u_2} \\
  \PD{ u^\ast, c \frac{\partial u}{\partial x} }
  &= \frac{c}{2\pi} \int_0^{2\pi} d\Theta \int_0^1 dy \,
  \biggl[ u^\ast \, u \biggr]_{x=0}^{x=2}
  + \PD{ -c \frac{\partial u^\ast}{\partial x} , u },
  \\
  \PD{ u^\ast, - \omega \frac{\partial u}{\partial \Theta} }
  &= -\frac{\omega}{2\pi} \int_0^2 dx \int_0^1 dy \,
  \biggl[ u^\ast \, u \biggr]_{\Theta=0}^{\Theta=2\pi}
  + \PD{ \omega \frac{\partial u^\ast}{\partial \Theta} , u }.
\end{align}
Using the Green's function $G(x, y, x', y')$ given in Eq.~(\ref{eq:G}),
the function $\psi_u(x - \Phi, y, \Theta)$ given in Eq.~(\ref{eq:P_u})
can also be written in the following form:
\begin{align}
  \psi_u(x - \Phi, y, \Theta)
  = \int_0^2 dx' \int_0^1 dy' \, G(x, y, x', y') \frac{\partial u(x' - \Phi, y', \Theta)}{\partial x'}.
\end{align}
In Eqs.~(\ref{eq:PD_P_u_1})(\ref{eq:PD_P_u_2}), we perform the following manipulations:
\begin{align}
  &\PD{ -\frac{\partial}{\partial x} \left[ u^\ast \left( \frac{\partial X_0}{\partial y} - 1 \right) \right] , \psi_u }
  \nonumber \\
  &= -\frac{1}{2\pi} \int_0^{2\pi} d\Theta \int_0^2 dx \int_0^1 dy \,
  \frac{\partial}{\partial x} \left[ u^\ast \left( \frac{\partial X_0}{\partial y} - 1 \right) \right] \, \psi_u
  \nonumber \\
  &= -\frac{1}{2\pi} \int_0^{2\pi} d\Theta \int_0^2 dx \int_0^1 dy \int_0^2 dx' \int_0^1 dy' \, G(x, y, x', y')
  \frac{\partial u'}{\partial x'}
  \frac{\partial}{\partial x} \left[ u^\ast \left( \frac{\partial X_0}{\partial y} - 1 \right) \right]
  \nonumber \\
  &= -\frac{1}{2\pi} \int_0^{2\pi} d\Theta \int_0^2 dx \int_0^1 dy \int_0^2 dx' \int_0^1 dy' \, G(x', y', x, y)
  \frac{\partial u}{\partial x}
  \frac{\partial}{\partial x'} \left[ {u^\ast}' \left( \frac{\partial X_0'}{\partial y'} - 1 \right) \right]
  \nonumber \\
  &= -\frac{1}{2\pi} \int_0^{2\pi} d\Theta \int_0^2 dx \int_0^1 dy \, \psi_{u, x}^\ast \frac{\partial u}{\partial x}
  \nonumber \\
  &= -\frac{1}{2\pi} \int_0^{2\pi} d\Theta \int_0^1 dy \,
  \biggl[ \psi_{u, x}^\ast \, u \biggr]_{x=0}^{x=2}
  + \PD{ \frac{\partial \psi_{u, x}^\ast}{\partial x} , u },
\end{align}
and
\begin{align}
  &\PD{ \frac{\partial}{\partial y} \left[ u^\ast \frac{\partial X_0}{\partial x} \right] , \psi_u }
  \nonumber \\
  &= \frac{1}{2\pi} \int_0^{2\pi} d\Theta \int_0^2 dx \int_0^1 dy \,
  \frac{\partial}{\partial y} \left[ u^\ast \frac{\partial X_0}{\partial x} \right] \, \psi_u
  \nonumber \\
  &= \frac{1}{2\pi} \int_0^{2\pi} d\Theta \int_0^2 dx \int_0^1 dy \int_0^2 dx' \int_0^1 dy' \, G(x, y, x', y')
  \frac{\partial u'}{\partial x'}
  \frac{\partial}{\partial y} \left[ u^\ast \frac{\partial X_0}{\partial x} \right]
  \nonumber \\
  &= \frac{1}{2\pi} \int_0^{2\pi} d\Theta \int_0^2 dx \int_0^1 dy \int_0^2 dx' \int_0^1 dy' \, G(x', y', x, y)
  \frac{\partial u}{\partial x}
  \frac{\partial}{\partial y'} \left[ {u^\ast}' \frac{\partial X_0'}{\partial x'} \right]
  \nonumber \\
  &= \frac{1}{2\pi} \int_0^{2\pi} d\Theta \int_0^2 dx \int_0^1 dy \, \psi_{u, y}^\ast \frac{\partial u}{\partial x}
  \nonumber \\
  &= \frac{1}{2\pi} \int_0^{2\pi} d\Theta \int_0^1 dy \,
  \biggl[ \psi_{u, y}^\ast \, u \biggr]_{x=0}^{x=2}
  + \PD{ -\frac{\partial \psi_{u, y}^\ast}{\partial x} , u },
\end{align}
where we used the following abbreviations:
\begin{align}
  X_0' = X_0(x' - \Phi, y', \Theta),
  \qquad
  u' = u(x' - \Phi, y', \Theta),
  \qquad
  {u^\ast}' = u^\ast(x' - \Phi, y', \Theta),
\end{align}
and also defined the following functions:
\begin{align}
  \psi_{u, x}^\ast(x - \Phi, y, \Theta)
  &= \int_0^2 dx' \int_0^1 dy' \, G(x', y', x, y)
  \frac{\partial}{\partial x'}
  \left[ {u^\ast}(x' - \Phi, y', \Theta) \left( \frac{\partial X_0(x' - \Phi, y', \Theta)}{\partial y'} - 1 \right) \right],
  \label{eq:Past_x} \\
  \psi_{u, y}^\ast(x - \Phi, y, \Theta)
  &= \int_0^2 dx' \int_0^1 dy' \, G(x', y', x, y)
  \frac{\partial}{\partial y'}
  \left[ {u^\ast}(x' - \Phi, y', \Theta) \frac{\partial X_0(x' - \Phi, y', \Theta)}{\partial x'} \right].
  \label{eq:Past_y}
\end{align}
Here, we note that Eqs.~(\ref{eq:Past_uast_x})(\ref{eq:Past_uast_y})
can be derived by applying the Laplacian to Eqs.~(\ref{eq:Past_x})(\ref{eq:Past_y}), respectively.
In this way, the adjoint operator ${\cal L}^\ast(x - \Phi, y, \Theta)$, defined in Eq.~(\ref{eq:operator}),
is obtained as
\begin{align}
  {\cal L}^\ast(x - \Phi, y, \Theta) u^\ast(x - \Phi, y, \Theta)
  = \frac{\partial^2 u^\ast}{\partial x^2}
  + \frac{\partial^2 u^\ast}{\partial y^2}
  + \frac{\partial}{\partial x} \left[ u^\ast \frac{\partial \psi_0}{\partial y} \right]
  - \frac{\partial}{\partial y} \left[ u^\ast \frac{\partial \psi_0}{\partial x} \right]
  + \frac{\partial \psi_{u,x}^\ast}{\partial x}
  - \frac{\partial \psi_{u,y}^\ast}{\partial x}
  - c \frac{\partial u^\ast}{\partial x}
  + \omega \frac{\partial u^\ast}{\partial \Theta}.
\end{align}
Similarly to the boundary conditions for $u(x - \Phi, y, \Theta)$ as
\begin{align}
  & u(x - \Phi + 2, y, \Theta) = u(x - \Phi, y, \Theta),
  \label{eq:bcux} \\
  & \Bigl. u(x - \Phi, y, \Theta) \Bigr|_{y = 0} = \Bigl. u(x - \Phi, y, \Theta) \Bigr|_{y = 1} = 0,
  \label{eq:bcuy} \\
  & u(x - \Phi, y, \Theta + 2\pi) = u(x - \Phi, y, \Theta),
  \label{eq:bcuT}
\end{align}
the adjoint boundary conditions are given by
\begin{align}
  & u^\ast(x - \Phi + 2, y, \Theta) = u^\ast(x - \Phi, y, \Theta),
  \label{eq:bcuastx} \\
  & \Bigl. u^\ast(x - \Phi, y, \Theta) \Bigr|_{y = 0} = \Bigl. u^\ast(x - \Phi, y, \Theta) \Bigr|_{y = 1} = 0,
  \label{eq:bcuasty} \\
  & u^\ast(x - \Phi, y, \Theta + 2\pi) = u^\ast(x - \Phi, y, \Theta),
  \label{eq:bcuastT}
\end{align}
which represent the periodic boundary condition on $x$,
the Dirichlet zero boundary condition on $y$,
and the $2\pi$-periodicity with respect to $\Theta$.
In fact, under these adjoint boundary conditions,
the bilinear concomitant
${\cal S}[ u^\ast(x - \Phi, y, \Theta), u(x - \Phi, y, \Theta) ]
= \pd{ u^\ast(x - \Phi, y, \Theta), {\cal L}(x - \Phi, y, \Theta) u(x - \Phi, y, \Theta) }
- \pd{ {\cal L}^\ast(x - \Phi, y, \Theta) u^\ast(x - \Phi, y, \Theta), u(x - \Phi, y, \Theta) }$
becomes zero, i.e.,
\begin{align}
  {\cal S}\Bigl[ u^\ast(x - \Phi, y, \Theta), u(x - \Phi, y, \Theta) \Bigr] =
  &+ \frac{1}{2\pi} \int_0^{2\pi} d\Theta \int_0^1 dy \,
  \left[ u^\ast \, \frac{\partial u}{\partial x} \right]_{x=0}^{x=2}
  \nonumber \\
  &- \frac{1}{2\pi} \int_0^{2\pi} d\Theta \int_0^1 dy \,
  \left[ \frac{\partial u^\ast}{\partial x} \, u \right]_{x=0}^{x=2}
  \nonumber \\
  &+ \frac{1}{2\pi} \int_0^{2\pi} d\Theta \int_0^2 dx \,
  \left[ u^\ast \, \frac{\partial u}{\partial y} \right]_{y=0}^{y=1}
  \nonumber \\
  &- \frac{1}{2\pi} \int_0^{2\pi} d\Theta \int_0^2 dx \,
  \left[ \frac{\partial u^\ast}{\partial y} \, u \right]_{y=0}^{y=1}
  \nonumber \\
  &-\frac{1}{2\pi} \int_0^{2\pi} d\Theta \int_0^1 dy \,
  \left[ u^\ast \, \frac{\partial \psi_0}{\partial y} \, u \right]_{x=0}^{x=2}
  \nonumber \\
  &+ \frac{1}{2\pi} \int_0^{2\pi} d\Theta \int_0^2 dx \,
  \left[ u^\ast \, \frac{\partial \psi_0}{\partial x} \, u \right]_{y=0}^{y=1}
  \nonumber \\
  &+ \frac{1}{2\pi} \int_0^{2\pi} d\Theta \int_0^1 dy \,
  \left[ u^\ast \, \left( \frac{\partial X_0}{\partial y} - 1 \right) \, \psi_u \right]_{x=0}^{x=2}
  \nonumber \\
  &- \frac{1}{2\pi} \int_0^{2\pi} d\Theta \int_0^2 dx \,
  \left[ u^\ast \, \frac{\partial X_0}{\partial x} \, \psi_u \right]_{y=0}^{y=1}
  \nonumber \\
  &- \frac{1}{2\pi} \int_0^{2\pi} d\Theta \int_0^1 dy \,
  \biggl[ \psi_{u, x}^\ast \, u \biggr]_{x=0}^{x=2}
  \nonumber \\
  &+ \frac{1}{2\pi} \int_0^{2\pi} d\Theta \int_0^1 dy \,
  \biggl[ \psi_{u, y}^\ast \, u \biggr]_{x=0}^{x=2}
  \nonumber \\
  &+ \frac{c}{2\pi} \int_0^{2\pi} d\Theta \int_0^1 dy \,
  \biggl[ u^\ast \, u \biggr]_{x=0}^{x=2}
  \nonumber \\
  &- \frac{\omega}{2\pi} \int_0^2 dx \int_0^1 dy \,
  \biggl[ u^\ast \, u \biggr]_{\Theta=0}^{\Theta=2\pi}
  = 0.
  \label{eq:calS}
\end{align}
Each term of the bilinear concomitant ${\cal S}[ u^\ast(x - \Phi, y, \Theta), u(x - \Phi, y, \Theta) ]$
vanishes for the following reasons:
the 1st, 2nd, 5th, 7th, 9th, 10th, and 11th terms in Eq.~(\ref{eq:calS}) become zero owing to
the $2$-periodicity with respect to $x$ for all the functions;
the 3rd, 4th, 6th, and 8th terms,
the Dirichlet zero boundary condition on $y$ for $u$ and $u^\ast$;
the last term (i.e., the 12th term),
the $2\pi$-periodicity with respect to $\Theta$ for all the functions.

\section{Oscillatory convection with weak boundary forcing} \label{sec:B}

In this appendix,
we consider oscillatory cylindrical-Hele-Shaw convection with weak boundary forcing
described by the following equation:
\begin{align}
  \frac{\partial}{\partial t} T(x, y, t) = \nabla^2 T + J(\psi, T),
  \label{eq:Tb}
\end{align}
where the stream function $\psi(x, y, t)$ is determined from the temperature field $T(x, y, t)$ as
\begin{align}
  \nabla^2 \psi(x, y, t) = -{\rm Ra} \frac{\partial T}{\partial x}.
  \label{eq:P_Tb}
\end{align}
The above two equations are a reproduction of Eqs.~(\ref{eq:T})(\ref{eq:P_T}) for readability.
The boundary conditions for the temperature field $T(x, y, t)$ are now given by
\begin{align}
  & \Bigl. T(x, y, t) \Bigr|_{y = 0} = 1 + \epsilon p_{\rm B}(x, t),
  \label{eq:bcTb} \\
  & \Bigl. T(x, y, t) \Bigr|_{y = 1} = 0 + \epsilon p_{\rm T}(x, t),
  \label{eq:bcTt}
\end{align}
where the weak boundary forcing applied at the bottom ($y = 0$) and the top ($y = 1$)
is described by $\epsilon p_{\rm B}(x, t)$ and $\epsilon p_{\rm T}(x, t)$, respectively.
The stream function $\psi(x, y, t)$ satisfies the Dirichlet zero boundary condition on $y$, i.e.,
\begin{align}
  \Bigl. \psi(x, y, t) \Bigr|_{y = 0} = \Bigl. \psi(x, y, t) \Bigr|_{y = 1} = 0.
\end{align}

Extending the transformation given by Eq.~(\ref{eq:T-X}),
we consider the following transformation of the temperature field
(see also, e.g., Ref.~\cite{ref:haberman12} for this type of transformation):
\begin{align}
  T(x, y, t) = (1 - y) + \epsilon P(x, y, t) + X(x, y, t),
  \label{eq:T-XwP}
\end{align}
where the function $P(x, y, t)$ is defined as
\begin{align}
  P(x, y, t) = p_{\rm B}(x, t) + y \Bigl[ p_{\rm T}(x, t) - p_{\rm B}(x, t) \Bigr].
  \label{eq:pTpB}
\end{align}
We note that
$P(x, y, t)|_{y = 0} = p_{\rm B}(x, t)$
and
$P(x, y, t)|_{y = 1} = p_{\rm T}(x, t)$.
By considering the boundary forcing,
the stream function $\psi(x, y, t)$ is also decomposed as
\begin{align}
  \psi(x, y, t) = \psi_{X}(x, y, t) + \epsilon \psi_{P}(x, y, t).
  \label{eq:P-PxPp}
\end{align}
Substituting Eqs.~(\ref{eq:T-XwP})(\ref{eq:P-PxPp}) into Eq.~(\ref{eq:Tb}),
we derive the following equation:
\begin{align}
  \frac{\partial}{\partial t} X(x, y, t)
  = \nabla^2 X + J(\psi_X, X) - \frac{\partial \psi_X}{\partial x}
  + \epsilon B(x, y, t) + \epsilon^2 J(\psi_P, P),
  \label{eq:XB}
\end{align}
where the first-order terms associated with the boundary forcing are given by
\begin{align}
  B(x, y, t)
  = \left[ \nabla^2 - \frac{\partial}{\partial t} \right] P
  + J(\psi_X, P) + J(\psi_P, X) - \frac{\partial \psi_P}{\partial x}.
\end{align}
Applying Eqs.~(\ref{eq:T-XwP})(\ref{eq:P-PxPp}) to Eq.~(\ref{eq:P_Tb})
and also considering the linearity of Eq.~(\ref{eq:P_Tb}),
we obtain the following equations:
\begin{align}
  \nabla^2 \psi_X(x, y, t) &= -{\rm Ra} \frac{\partial X}{\partial x},
  \label{eq:Px_X} \\
  \nabla^2 \psi_P(x, y, t) &= -{\rm Ra} \frac{\partial P}{\partial x}.
  \label{eq:Pp_P}
\end{align}
We note that Eq.~(\ref{eq:Px_X}) corresponds to Eq.~(\ref{eq:P_X}).
From Eqs.~(\ref{eq:bcTb})(\ref{eq:bcTt}) and Eqs.~(\ref{eq:T-XwP})(\ref{eq:pTpB}),
as in Eq.~(\ref{eq:bcXy}),
the convective component $X(x, y, t)$ satisfies the Dirichlet zero boundary condition on $y$, i.e.,
\begin{align}
  \Bigl. X(x, y, t) \Bigr|_{y = 0} = \Bigl. X(x, y, t) \Bigr|_{y = 1} = 0.
\end{align}
The two stream functions, $\psi_X(x, y, t)$ and $\psi_P(x, y, t)$,
also satisfy the Dirichlet zero boundary condition on $y$, i.e.,
\begin{align}
  \Bigl. \psi_X(x, y, t) \Bigr|_{y = 0} = \Bigl. \psi_X(x, y, t) \Bigr|_{y = 1} &= 0,
  \\
  \Bigl. \psi_P(x, y, t) \Bigr|_{y = 0} = \Bigl. \psi_P(x, y, t) \Bigr|_{y = 1} &= 0.
\end{align}
In this way,
oscillatory convection with boundary forcing is found to be exactly described by Eq.~(\ref{eq:XB}),
which possesses the form of Eq.~(\ref{eq:Xp}) from the viewpoint of the phase reduction.

When the boundary forcing is absent, i.e., $\epsilon = 0$,
the system is assumed to exhibit oscillatory cylindrical-Hele-Shaw convection
described by the following limit-torus solution:
\begin{align}
  X(x, y, t) = X_0\bigl( x - \Phi(t), y, \Theta(t) \bigr), \qquad
  \dot{\Phi}(t) = c, \qquad
  \dot{\Theta}(t) = \omega,
\end{align}
which is a reproduction of Eq.~(\ref{eq:X_X0}) for readability.
Under the assumption that the boundary forcing is sufficiently weak,
as in Sec.~\ref{subsec:2E},
using the phase sensitivity functions given by Eqs.~(\ref{eq:Zs})(\ref{eq:Zt}),
we derive a set of phase equations from Eq.~(\ref{eq:XB}) as follows:
\begin{align}
  \dot{\Phi}(t)
  &= c + \epsilon \int_0^2 dx \int_0^1 dy \,
  Z_{\rm s}(x - \Phi, y, \Theta) B_0(x, x - \Phi, y, \Theta, t),
  \\
  \dot{\Theta}(t)
  &= \omega + \epsilon \int_0^2 dx \int_0^1 dy \,
  Z_{\rm t}(x - \Phi, y, \Theta) B_0(x, x - \Phi, y, \Theta, t),
\end{align}
where the {\it effective boundary forcing function} is given by
\begin{align}
  B_0(x, x - \Phi, y, \Theta, t)
  = \left[ \nabla^2 - \frac{\partial}{\partial t} \right] P
  + J(\psi_0, P) + J(\psi_P, X_0) - \frac{\partial \psi_P}{\partial x}.
\end{align}
We note that the two functions, $X_0$ and $P_0$, are given by
Eq.~(\ref{eq:X0}) and Eq.~(\ref{eq:P0}), respectively.
We also note that the second-order term of Eq.~(\ref{eq:XB}),
i.e., $\epsilon^2 J(\psi_P, P)$,
is negligible owing to the smallness of $\epsilon$.
Furthermore, we consider the case in which the weak forcing is spatially homogeneous as follows:
\begin{align}
  p_{\rm B}(x, t) &= q_{\rm B}(t),
  \\
  p_{\rm T}(x, t) &= q_{\rm T}(t).
\end{align}
For this case, the effective boundary forcing function is simplified as
\begin{align}
  B_0(x - \Phi, y, \Theta, t)
  = \Bigl[ q_{\rm T}(t) - q_{\rm B}(t) \Bigr] \frac{\partial \psi_0}{\partial x}
  - \Bigl[ y \, \dot{q}_{\rm T}(t) + (1 - y) \, \dot{q}_{\rm B}(t) \Bigr].
\end{align}

In summary, the phase description method is applicable to
the oscillatory cylindrical-Hele-Shaw convection with weak forcing
applied to the boundary given in Eqs.~(\ref{eq:bcTb})(\ref{eq:bcTt})
as well as to the bulk given in Eq.~(\ref{eq:Tp}).
Because the spatial extent of the cylindrical-Hele-Shaw cell is quasi-two-dimensional,
external forcing can be easily applied not only to the boundary but also to the bulk in experiments;
however, in general, although external forcing can be easily applied to boundary,
it may be difficult to apply external forcing to bulk of fluid systems in experiments.
In fact, in the rotating fluid annulus experiments~\cite{ref:read09,ref:read10},
perturbations are applied to the boundary.
Therefore, the treatments of boundary forcing as developed in this appendix
are required to apply the phase description method to such fluid systems.


\clearpage


\begin{figure*}
  \begin{center}
    \includegraphics[width=0.5\hsize,clip]{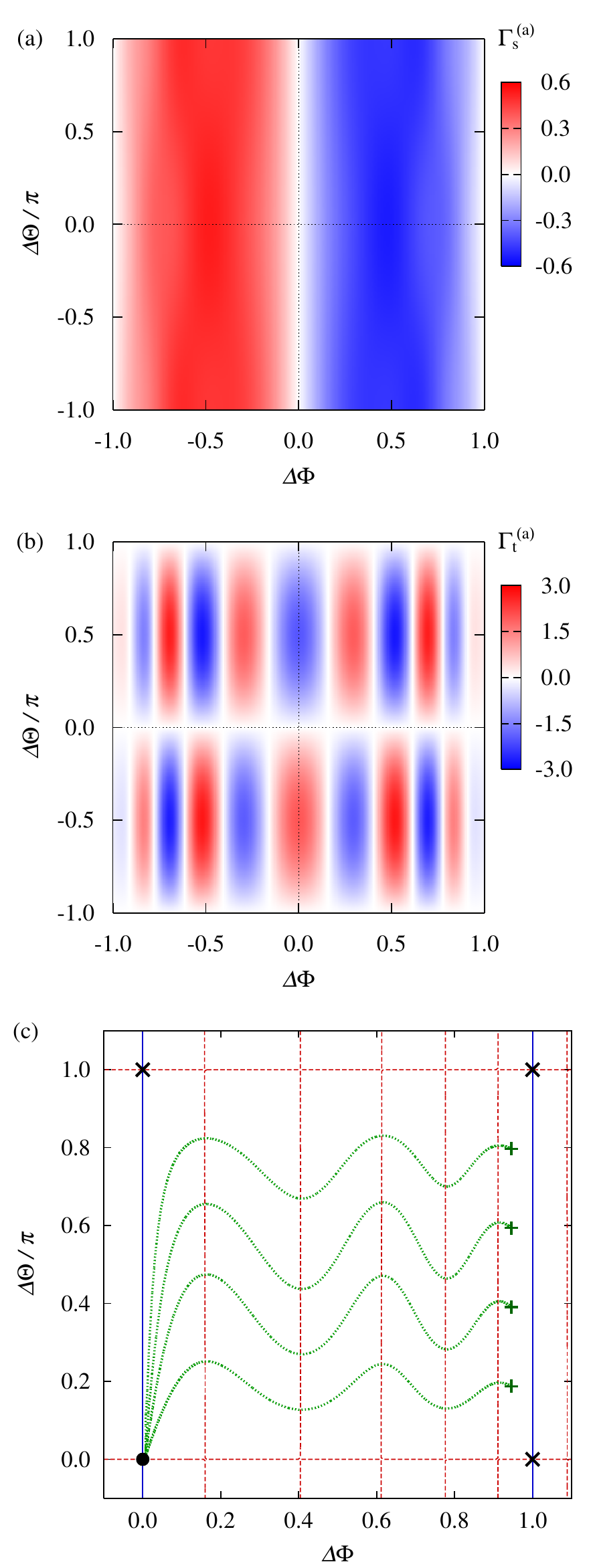}
  \end{center}
\end{figure*}
\clearpage
\begin{figure*}
  \begin{center}
  \end{center}
  \caption{(Color online)
    (a) Anti-symmetric component of the phase coupling function for the spatial phase,
    $\Gamma_{\rm s}^{\rm (a)}(\varDelta \Phi, \varDelta \Theta)$.
    (b) Anti-symmetric component of the phase coupling function for the temporal phase,
    $\Gamma_{\rm s}^{\rm (a)}(\varDelta \Phi, \varDelta \Theta)$.
    (c) Nullclines, fixed points, and typical orbits of the phase differences
    in the following region $(\varDelta \Phi, \varDelta \Theta / \pi) \in [0, 1] \times [0, 1]$.
    The solid (blue) and broken (red) lines indicate the nullclines of
    $\Gamma_{\rm s}^{\rm (a)}(\varDelta \Phi, \varDelta \Theta)$ and
    $\Gamma_{\rm t}^{\rm (a)}(\varDelta \Phi, \varDelta \Theta)$, respectively.
    The filled circle ($\bullet$) indicates the stable fixed point
    $(\varDelta \Phi, \varDelta \Theta / \pi) = (0, 0)$,
    whereas the times signs ($\times$) indicate the unstable fixed points
    $(\varDelta \Phi, \varDelta \Theta / \pi) = (0, 1)$, $(1, 0)$, and $(1, 1)$.
    The dotted (green) lines indicate the typical orbits of the phase differences
    whose initial values indicated by the plus signs ($+$) are
    $(\varDelta \Phi, \varDelta \Theta / \pi) = (0.95, 0.20)$,
    $(0.95, 0.40)$, $(0.95, 0.60)$, and $(0.95, 0.80)$.
  }
  \label{fig:12}
\end{figure*}

\clearpage

\end{document}